\documentclass[journal, 10pt, twocolumn, balance]{IEEEtran}
\IEEEoverridecommandlockouts    

\usepackage{stmaryrd}
\usepackage{amsfonts}

\usepackage{amssymb}
\usepackage{amsmath}
\usepackage{amsmath,bm}
\usepackage{mathrsfs}
\usepackage{verbatim}
\usepackage{multirow}
\usepackage{array}

\ifCLASSINFOpdf
	\usepackage[pdftex]{graphicx}
	\graphicspath{{./}}
	\DeclareGraphicsExtensions{.pdf,.jpeg,.png}
	\usepackage[pdftex, bookmarks = true, colorlinks=false]{hyperref}
\else
	\usepackage[dvips]{graphicx}
	\graphicspath{{./}}
	\DeclareGraphicsExtensions{.eps}
	\usepackage[dvips, bookmarks=true, colorlinks=false]{hyperref}
\fi
\RequirePackage{lineno}

\usepackage{color}
\usepackage[usenames]{xcolor}

\usepackage{arydshln,leftidx,mathtools}
\usepackage{booktabs}
\usepackage{subfig}

\DeclareMathOperator*{\argmax}{arg\,max}

\begin{document}

\setlength{\dashlinegap}{1.0pt}



\title{A Three-State Received Signal Strength Model \\ for Device-free Localization
\thanks{Ossi Kaltiokallio, H\"{u}seyin~Yi\u{g}itler, and Riku~J\"{a}ntti are with the Department of Communications and Networking, Aalto University, School of Electrical Engineering, Espoo, Finland (email:\{name.surname\}@aalto.fi).}}

\author{Ossi Kaltiokallio, H\"{u}seyin~Yi\u{g}itler and Riku~J\"{a}ntti}

\maketitle

\begin{abstract}
The indoor radio propagation channel is typically modeled as a two-state time-variant process where one of the states represents the channel when the environment is static, whereas the other state characterizes the medium when it is altered by people. In this paper, the aforementioned process is augmented with an additional state. It is shown that the changes in received signal strength are dictated by: i) electronic noise, when a person is not present in the monitored area; ii) reflection, when a person is moving in the close vicinity of the line-of-sight; iii) shadowing, when a person is obstructing the line-of-sight component of the transmitter-receiver pair. Statistical and spatial models for the three states are derived and the models are empirically validated. Based on the models, a simplistic device-free localization application is designed which aims to: first, estimate the temporal state of the channel using a hidden Markov model; second, track a person using a particle filter. The results suggest that the tracking accuracy is enhanced by at least 65\% while the link's sensitivity region is increased by 100\% or more with respect to empirical models presented in earlier works.
\end{abstract}

\begin{IEEEkeywords}
received signal strength, indoor radio propagation channel, temporal fading, device-free localization
\end{IEEEkeywords}

\section{Introduction} \label{S:introduction}
\begin{table*}
    \caption{Major notations} 
        \centering 
	\renewcommand{\arraystretch}{1.2}
        \begin{tabular}{| l | c | l |} 
        
	\hline\hline
        \multirow{2}{*}{Appearance} & \multirow{2}{*}{Parameters} & \multirow{2}{*}{Description (the order of description is the same as the order of appearance in parameters column)}  \\
	& & \\ \hline
	
	\multirow{2}{*}{\emph{RF Signal}}  & $\alpha(t), \tau(t), \phi(t)$& Time-variant amplitude, time-delay and phase of the RF signal. \\
	& $f_c, \lambda_c$ & Time-invariant carrier frequency and wave length of the transmitted low-pass signal. \\ \hline
	
	\emph{Measurement}  & $h(k), \hat{R}(k) $   & Channel gain and RSS output of the RX at time instant $k$. \\
	\emph{Model} & $ r(k), g(k), \nu(k)$ & Mean-removed and pre-filtered RSS, the signal of interest and wideband noise of the measurements.  \\ \hline
	
	\emph{Statistical and}  & $\mathcal{R}, \mathcal{S}$ & Reflection and shadowing models. \\ 
	\emph{Spatial Models} & $f_{r|s}$ & Density of $r(k)$ conditioned on temporal state of the propagation channel. \\ \hline
	
	\multirow{2}{*}{\emph{Human Ellipse}}  & $\bm{p_c}, A, B$ & Center coordinates of the person, and semi-minor and major axis of the human ellipse model.\\
	\multirow{2}{*}{\emph{Model}}  &  $\varepsilon_r, \rho$  & Relative permittivity and attenuation factor of the person. \\
	
	 & $\Delta, \Delta_R$ & Excess path length with respect to $\bm{p_c}$ and reflection point $\bm{p_R}$. \\ \hline

	\multirow{2}{*}{\emph{Application}}  & $s, \hat{s}$  &Temporal state of the propagation channel and its estimate.   \\ 
	& $\bm{x}, \bm{\hat{x}}$ &  Kinematic state of the person and its estimate. \\ \hline
	
	\multirow{2}{*}{\emph{Experiments}}  & $T_s, c, C$  & Sampling interval, frequency channel and number of used frequency channels.  \\
	& $\bm{p}_{TX}, \bm{p}_{RX}, d_{LoS}$ &  Coordinates of the TX, RX and distance between them.\\ \hline
	
	\multirow{2}{*}{\emph{Evaluation}}  & $\bar{\epsilon}_x, \bar{\epsilon}_y$  & Mean absolute error of coordinate estimates.  \\
	& $\bar{\epsilon}_\%, \bar{\epsilon}_R$ &  Ratio of particles within modeled human ellipse and the enhancement in accuracy.\\ \hline
	
        \end{tabular}
        \label{table:major_notation} 
\end{table*}

\PARstart{T}{he success} of wireless communication systems together with the recent advances in different technologies have enabled the development of wireless sensor networks (WSNs) \cite{Akyildiz2002}. These networks are typically composed of inexpensive nodes which are limited by processing power, memory, bandwidth, communication range, and available energy. Despite the limitations, WSNs also possess several advantages over their cabled counterparts such as allowing development of reconfigurable and expandable autonomous systems, enabling use-case scenarios which are impossible for cabled systems, in addition to reducing deployment time and complexity. Therefore, WSNs have attracted considerable attention in the research community and currently they are used and tested for example in: health care \cite{Ko2010b}, wireless control \cite{kaltiokallio2011}, and structural health monitoring \cite{Bocca2011}. In addition to the more conventional application areas, low-cost of the transceivers enable dense network deployments rendering new sensing possibilities such as: device-free localization (DFL) \cite{patwari08}, fall detection \cite{mager2013}, and non-invasive breathing rate monitoring \cite{ Patwari2011}. These systems exploit measurements of the radio to extract information about people in the surrounding environment and therefore, such networks are referred to as RF sensor networks \cite{patwari2010}. Most notably, these networks do not require people to co-operate with the system, allowing one to gain situational awareness of the environment non-invasively. 
	
Typically, RF sensor network applications use the received signal strength (RSS) measurements of low-cost and static transceivers. For such applications, the problem at hand is three-fold. First, the temporal RSS variations need to be accurately modeled to identify when and how a link is affected by people. Second, the RSS has to be related to spatial information of the transceivers to enable localization of people. Third, the RSS is a measure of the average received signal power resulting to the fact that individual signals altered by a specific object cannot be resolved. Thus, localization methods developed for radar and ultra-wideband systems cannot be applied. This paper focuses on the first two items listed above with the aim to statistically and spatially model the temporal RSS changes with respect to the location of a person. The developed models are demonstrated with a DFL application.

Electromagnetic waves can be reflected, diffracted, scattered and attenuated as they propagate from TX to RX. The dominating mechanism observed at the RX depends on: properties of the transmitted signal, the signals interaction with intervening objects, and location of the RX with respect to the TX \cite[Ch. 4]{Molisch2010}. For wavelengths smaller than the geometrical extent of a human body, it has already been shown that shadowing dictates the RSS measurements when a person obstructs the line-of-sight (LoS) component of the TX-RX pair \cite{patwari08} whereas reflection, when a person is in the close vicinity of the LoS \cite{patwari2011b}. Even though it has been identified how people affect RF signals, to the best of our knowledge, we are the first in the DFL society to propose a framework to identify the dominating propagation mechanism and adjust the measurement models accordingly.

Typically, the number and location of intervening objects and their effect on individual multipath components are unknown. Therefore, the indoor propagation channel is usually modeled as a stochastic process \cite[Ch. 5]{Molisch2010} and considerable efforts in the late $80$'s and early $90$'s were made to characterize the channel to enable successful wireless communication system deployments \cite{hashemi93}. Among other findings, it was identified that RF signals experience time-intervals of considerable fading caused by the movements of people, whereas most of the time the channel remains nearly constant \cite{bultitude1987, hashemi94}. This observation motivated to model the time-varying process using a two-state Markov chain \cite{roberts1995}, a result exploited in DFL applications to identify when a wireless link is affected by people and when it is not \cite{Wilson2012, Zheng2012}. Further, it is a common practice in the DFL society to suppose that shadowing is the only source of fading and that a person is very close to or obstructing the LoS at these time instances \cite{Wilson2010,li2011,Kaltiokallio2012}. However, this approach does not include the cases where fading is caused by reflections. In this work, we address this issue by augmenting the aforementioned two-state temporal model with an additional state so that both reflections and shadowing are accounted for. As an outcome, the localization accuracy of the system is improved while the spatial extent of a link's sensing region is increased.

In this paper, it is demonstrated that the temporal RSS changes are caused by: i) electronic noise, when a person is not present in the monitored area; ii) reflection, when a person is moving in the close proximity of the LoS; iii) shadowing, when a person is obstructing the LoS. A three-state model for the temporal RSS changes is proposed and statistical and spatial models for the different states are derived. The introduced models are demonstrated using a DFL application where the system estimates: first, the temporal state of the channel using a hidden Markov model (HMM) and second, coordinates of the person using a particle filter. With the models proposed in this paper, the results suggest an enhancement in tracking accuracy by $65 - 350 \%$ whereas an increase in the link's sensing region by $100 \%$ or more with respect to empirical models proposed in earlier literature.

The rest of the paper is organized as follows. In Section \ref{S:related_work}, related work is discussed. In Section \ref{S:model}, a three-state RF propagation model is presented and in Section \ref{S:spatial_models}, statistical and spatial models for the different temporal states are derived. The DFL application and the conducted experiments are introduced in Sections \ref{S:application} and \ref{S:experiments} respectively. Results are presented in Section \ref{S:results} and conclusions are drawn in \ref{S:conlcusions}. In Table~\ref{table:major_notation}, major notations of this paper are summarized.

\section{Related Work} \label{S:related_work}
The indoor propagation channel has been studied comprehensively e.g. by Hashemi \cite{hashemi93}, Bultitude \cite{bultitude1987} and Wyne \emph{et al.} \cite{Wyne2009} and regarding this paper, some of the key findings are: i) the channel is non-stationary in time; ii) the impulse response profiles for points in close vicinity of one another are correlated; iii) adjacent multipath components of the same impulse response are not independent; iv) the channel's parameters strongly depend on the measurement setup and environment. Considering (i), it is widely accepted that fading occurs in bursts \cite{bultitude1987} and this fading/non-fading time-varying process has motivated the use of a two-state Markov model \cite{roberts1995} where the states are characterized as Ricean variates with different $K$-factors. However, the Ricean distribution is based on the central limit theorem and it assumes that the scatterer locations are random violating (ii) and (iii) above. Supporting (ii) and (iii), the arrival times of different multipath components are correlated \cite{hashemi93}. Further, in case there are dominating scattering planes (e.g. floor and walls), the Ricean distribution (even though an appropriate functional fit) is not a valid theoretical model for the received signal envelope \cite{Wyne2009}. Despite its wide acceptance, the Ricean distribution has its own shortcomings and therefore, we investigate different statistical models to characterize the indoor propagation channel in LoS environments.

People can influence RF signals in various ways including reflection, diffraction, scattering and attenuation \cite[Ch. 4]{Molisch2010}. Of the different mechanisms, it has been shown that shadowing dictates the RSS when a person obstructs the LoS \cite{patwari08} whereas reflection, when a person is in the close vicinity of the LoS \cite{patwari2011b}. On the other hand, the effects of diffraction and scattering are typically neglected in DFL since the former has a complex relationship to the object’s geometry \cite{patwari2011b} whereas the latter is not expected to contribute to the RSS significantly. Therefore, we present a temporal model which consists of three states contrary to two-state models presented in earlier works \cite{Wilson2012, Zheng2012}. The considered states are: \emph{non-fading}, \emph{reflection} and \emph{shadowing}, and to the best of our knowledge, we are the first to propose a measurement modality which includes both reflection and shadowing mechanisms. The derived deterministic model for reflection is based on a \emph{single-bounce} model \cite[pp. 114-125]{rappaport1996}, whereas the shadowing model is based on the theory of \emph{computerized tomographic imaging} (CTI) \cite[Ch. 3]{Kak1988}. 

The authors wish to emphasize at this point that the geometric extent of the human body must be considered in the design process when deriving deterministic models. Human geometry is typically neglected in related works by modeling the person as a point in the 2D Cartesian coordinate system \cite{li2011,Zheng2012,Guo2013}. Further, high sampling rate is of great importance in the design process since it enables to observe even small changes in the propagation channel. In DFL deployments, the coherence time of the channel is  typically not considered when deciding on sampling frequency of the system \cite{Wilson2012, li2011, Kaltiokallio2012, Kaltiokallio2013}. Thus, the RSS measurements are not altered by the same propagation medium and therefore, achievable accuracy of these systems is limited.

\section{A Three-state RSS Measurement Model} \label{S:model}
\subsection{Preliminaries}

In this paper, the effect of people to RSS of narrow-band wireless communication devices is investigated. The development efforts focus, but do not limit, to coherent receiver architectures since for such devices it is straight forward to draw the relation between temporal variations in the propagation channel with respect to changes in RSS. The considered wireless devices operate at the $2.4 \text{ GHz}$ ISM-band which is a suitable frequency band to study human-induced RSS changes. For $2.4 \text{ GHz}$, the geometrical extent of a human body is considerably larger than the wavelength and therefore, it is expected that a person has a significant effect on the measured RSS. Further, we assume that the RX synchronizes to the LoS component so that the changes in RSS are with respect to the LoS signal.

We consider the scenarios where a single transmitter and one or more receivers are deployed in the environment. During operation, a single person is moving in the close vicinity of the LoS; otherwise, the propagation medium is assumed to be stationary. The used wireless devices enable communication over multiples of closely spaced frequency channels and we exploit channel diversity to enrich the low-resolution RSS measurements as in \cite{Kaltiokallio2012a}. We also assume that the fading process is slow so that the channel can be oversampled. Further, the coherence bandwidth of the channel is presumed to span the entire communication bandwidth. Thus, we expect to capture even small changes in the channel and that the changes are observable on all frequencies.

\subsection{Measurement Model}

RF waves emitted by a narrow-band communication system operating at carrier frequency $f_c$ are altered by the propagation channel so that the low-pass equivalent of the received signal \cite[Ch. 13]{Proakis2008} can be written as  
\begin{equation}\label{eq:received_signal}
	\zeta(t) = \sum_{i} \alpha_i(t) e^{-j\phi_i(t)}s(t-\tau_i(t)),
\end{equation}
where $s(t)$ is the transmitted low-pass signal and $\alpha_i(t)$, $\tau_i(t)$, $\phi_i(t)= 2 \pi f_c \tau_i(t)$ are the time-varying amplitude, time-delay, and phase of the $i^ \text{th}$ multipath component in respective order. The RSS is a measure of the received signal power and it is calculated over several periods of $s(t)$. If the channel is stationary for this duration, the RSS can be expressed as
\begin{equation}\label{eq:narrowband}
	R_{mW}(t) = P_0 \biggl| \sum_{i} \alpha_i(t) e^{-j\phi_i(t)}\biggr|^2,
\end{equation} 
where $P_0$ is a communication system dependent gain, which accounts for the transmitted signal power, properties of TX and RX electronics, and antenna gains. In logarithmic scale, the RSS is given by
\begin{equation}\label{eq:RSS_dB}
	R_{dB}(t) = P_{dB} + h_{dB}(t),
\end{equation}
where $P_{dB} = 10\log_{10}(P_0)$ and $h_{dB}(t)$ is the channel gain and its linear scale equivalent is
\begin{equation}\label{eq:channel}
	h(t) = \biggl|\sum_{i} \alpha_i(t) e^{-j\phi_i(t)}\biggr|^2.
\end{equation}

A typical narrow-band receiver samples and outputs the RSS every $T_s$ seconds, so that the measurement at time $t = k T_s$, $k \in \mathbb{N},$ is given by
\begin{equation}
	\hat{R}_{dB}(k) = P_{dB} + h_{dB}(k) + \hat{\nu}_{dB}(k),
\end{equation}
where $\hat{\nu}_{dB}(k)$ is zero-mean additive wideband noise. If the channel is quasi-static for the duration of interest, the channel is mean ergodic so that the statistical expectation $\mathrm{E}\{\cdot\}$ of the measurements is equivalent to the time average. The expected value of RSS under static channel conditions, given by
\begin{equation}
\mathrm{E}\{\hat{R}_{dB}(k)\}= P_{dB} + \mathrm{E}\{h_{dB}(k)\},
\end{equation}
reflects the site dependent and time-invariant channel characteristics. Thus, removing the mean yields a signal which does not depend on the underlying measurement setup and environment. 

Time variations of the mean removed RSS, given by
\begin{equation}\label{eq:rss_mean_removed_filtered}
\begin{aligned}
\tilde{R}_{dB}(k) &= \hat{R}_{dB}(k) - \mathrm{E}\{\hat{R}_{dB}(k)\} \\ 
	&= h_{dB}(k) - \mathrm{E}\{h_{dB}(k)\} + \hat{\nu}_{dB}(k),
\end{aligned}
\end{equation}
reflects the possible changes in the propagation channel. Further, the rate at which the propagation medium is being altered e.g. by a moving person dictates the frequency content of $\{\tilde{R}_{dB}(k)\}_{k=1}^K$; a phenomena referred as the Doppler spectra. Considering wireless devices of today, it is not hard to communicate or in our case sample the channel at a rate much higher than the maximum Doppler spread. In this paper, the channel is over-sampled and $\tilde{R}_{dB}(k)$ is low-pass filtered to increase the signal-to-noise ratio (SNR) of the measurements. We denote the mean-removed and filtered RSS measurements on frequency channel $c$ as
\begin{equation} \label{eq:rss}
    r_c(k) =  g_c(k) + \nu_c(k),
\end{equation}
where $g_c(k) = h_{dB}(k) - \mathrm{E}\{h_{dB}(k)\}$ and $\nu_c(k) $ is assumed to have a flat spectra.

Most wireless devices enable communication over different frequency channels $c$ so that Eq.~\eqref{eq:rss} can be extended to a measurement vector 
\begin{equation}
	\bm{r}(k) = \left[ \begin{array}{l l c l} r_1(k)&r_2(k)&\cdots&r_C(k) \end{array} \right]^T,
\end{equation}
where $C$ is the number of used channels. Consequently, the measurement model of the studied system is given by
\begin{equation}  \label{eq:measurementmodel}
	\boldsymbol{r}(k) = \boldsymbol{g}(k) + \boldsymbol{\nu}(k), 
\end{equation}
where $\bm{\nu}(k)$ is the filtered noise vector, which is assumed to be a zero-mean multivariate Gaussian of independent components with equal variances $\sigma^2$, i.e., $ \bm{\nu}(k) \sim \mathcal{N}(\bm{0}, \sigma^2 \bm{I})$.

For the considered scenario, $g_c(k)$ is a function of position, geometry and electrical properties of the person. Thus, if fading is parametrized with respect to these parameters by assuming them as deterministic variables, $g_c(k)$ is also deterministic. On the other hand, if some or all parameters are not precisely known, then $g_c(k)$ is stochastic. The deterministic models are the most useful when the RSS is calculated using the known parameters for example to examine the \emph{innovation} in a given measurement. The stochastic models on the other hand can be utilized to investigate the expected performance or to calculate the probability of measuring the acquired RSS. Therefore, depending on the perspective, both deterministic and stochastic models of $g_c(k)$ are useful and important.

\subsection{Temporal Characteristics of Human-induced RSS Changes}\label{S:temporal_characteristics}

A typical receiver is only sensitive to multipath components that arrive when $\tau_i < T$, where time constant $T$ is defined by parameters of the communication system such as: transmission power, receiver sensitivity, variance of electronic noise, TX-RX distance, and other deployment parameters. Since $\tau_i$ is a function of traversed distance, $T$ defines the \emph{sensitivity region} of the TX-RX pair which can be represented as an ellipse having the transceivers located at the foci \cite{Chang2004}. Thus, the RSS of a static TX-RX pair is only affected by changes within the sensitivity region.

Suppose an ideal scenario where for all static scattering planes in the environment we have $\tau_i > T$ so that the LoS component is the only one received. If a person is moving outside the sensing region, the propagation channel is stationary and $\nu_c$ dictates the statistics of $r_c$. Since the mean is removed, $g_c$ is expected to be zero because a single realization of the fading process is observed.

If the person now starts to move along a path perpendicular and toward the LoS, at some point, they enter the sensitivity region of the transceivers. After this, human-induced temporal fading can be observed at the RX since the person is creating additional multipath components with $\tau_i < T$. At these time instances, it is expected that reflection is the dominating propagation mechanism \cite{patwari2011b}. These reflected waves cause constructive and destructive fading sequentially because the person travels through odd and even Fresnel zones. Moreover, as the person moves closer to the LoS, incidence angle of reflected signal decreases, increasing amplitude of the reflected signal and therefore, altering the RSS more and more \cite[pp. 114-125]{rappaport1996}.

The considered process changes considerably as the person finally obstructs the LoS component of the TX-RX pair. Now, the RF signal(s) traverse through and around the human as they reflect, diffract, scatter and attenuate upon contact to non-homogeneous human tissue. At this state, $g_c$ is expected to be a stochastic signal since various propagation mechanism are affecting the measurements $-$ possibly simultaneously.  However, it is widely accepted that human's shadow the link at this state since typically a large decrease in signal strength is observed \cite{patwari08, li2011, Guo2013}. With considerable simplifications about geometry and electrical properties of the person, $g_c$ can be modeled as a deterministic signal. In this case, transmission through the human body is considered as the dominating propagation mechanism and the decrease in RSS can be modeled with respect to distance that the wave travels inside the body.

\subsection {Three-State Temporal Model}

The temporal state of the propagation channel defines the variations in RSS as discussed in Section~\ref{S:temporal_characteristics}. Therefore, in this paper $g_c(k)$ in Eq.~\eqref{eq:rss} is represented with respect to the state of the channel $s$ as
\begin{equation}\label{eq:signal}
    g_c(k;s) =  \begin{cases}
			       0 & s = s_1 \text{ (\emph{non-fading})}, \\
	\mathcal{R}(k, \bm{p_c}) & s = s_2 \text{ (\emph{reflection})}, \\
	\mathcal{S}(k, \bm{p_c}) & s = s_3 \text{ (\emph{shadowing})}. \\
               \end{cases}
\end{equation}
In Eq.~\eqref{eq:signal}, $\mathcal{R}(k, \bm{p_c})$ and $\mathcal{S}(k, \bm{p_c})$ are models for reflection and shadowing in respective order and they are functions of location $\bm{p_c} = (p_x, p_y)$, geometry and electrical properties of the human body.

In order to develop the deterministic models for reflection and shadowing, we approximate the cross section of a person with an ellipse that has uniform electrical properties. These simplifications allow us to derive a closed form expression for both $\mathcal{R}(k, \bm{p})$ and $\mathcal{S}(k, \bm{p})$. On the other hand, the stochastic models for $r_c$ follow the joint statistics of $g_c$ and $\nu_c$. However, it is expected that the statistics of $g_c$ dominate $\nu_c$ in case fading occurs and the density of $r_c$ conditioned on $s$ is anticipated to follow the statistics of $g_c$.

\section{Statistical and Spatial Models} \label{S:spatial_models}
\subsection{ Measurement Collection}\label{sec:measurements_collection}

\begin{figure*}[!t]
\begin{centering}
\begin{tabular}{*2{>{\centering\arraybackslash}m{0.95\columnwidth}}}
\subfloat[Single-bounce reflection model]{\includegraphics[width=0.95\columnwidth]{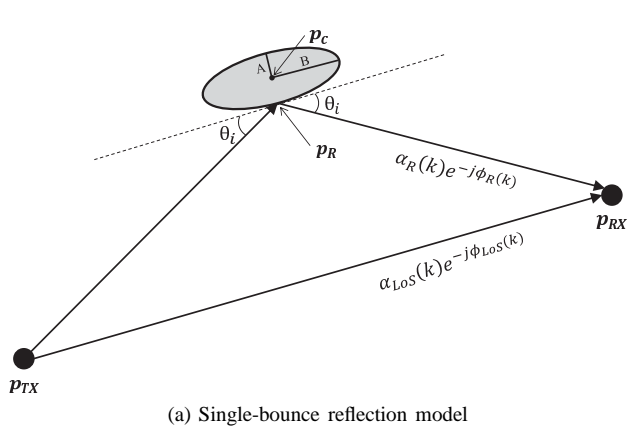}\label{fig:reflections}} &
\subfloat[$r(k)$ vs. $\mathcal{R}(k, \boldsymbol{p_c})$ given in Eq.~\eqref{eq:reflection_model}]{\includegraphics[width=\columnwidth]{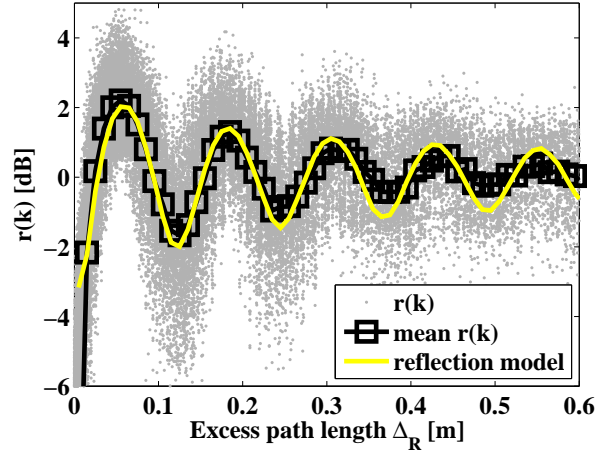}\label{fig:reflection_model}}
\end{tabular}
\caption{Model for human-induced reflections. In (b), $\psi_0 = 0.5$, $\varepsilon_r = 1.5$ and $\eta = 2$} 
\label{fig:reflection}
\end{centering}
\end{figure*}

The models introduced in this section are validated using RSS measurements of a single TX-RX pair deployed in an open indoor environment on podiums at a height of $1.0$ m, $3.0$ m apart from each other. Both of the nodes are equipped with Texas Instruments CC2431 IEEE 802.15.4 PHY/MAC compliant $2.4 \text{ GHz}$ transceivers \cite{CC2430} where the transceiver's micro-controller units run a communication software and a modified version of the FreeRTOS micro-kernel operating system \cite{Freertos}. In the experiments, directional antennas are used to assure that the RX synchronizes to the LoS component and that the changes in RSS only reflect variations in the close vicinity of the LoS. The used directional antennas provide a $8$ dBi gain and a horizontal beam width of $75^{\circ}$ \cite{lcom}.

The TX is programmed to transmit packets over each of the $16$ frequency channels defined by the IEEE 802.15.4 standard \cite{802_15_4} and after each transmission, the TX changes the frequency channel of communication in sequential order. The RX is programmed to receive the packets and to store the data to a SD card for offline analysis. In the experiments, oversampling is exploited and on average, the reception interval between two consecutive receptions is $2\text{ ms}$ with a standard deviation of $141 \text{ microseconds}$ resulting that the sampling interval $T_s$ of each frequency channel is $32 \text{ ms}$. 

In this paper, we only consider the scenario where a single person is within the sensing region of the nodes. In the experiments, markers are placed on the floor for the person to follow and a metronome is used to set a pre-defined walking pace. The person is walking at a speed of $0.5 \text{ m/s}$ intersecting the LoS multiple times. The experiment is repeated five times and in between each test, the RX is moved fourth of a wave length $(\lambda/4 \approx 3 \text{ cm})$ further away from the TX to include small spatial variance to the measurements. In total, $8.72 \cdot 10^5$ RSS measurements are collected and each measurement is associated to the true location of the person. A finite impulse response (FIR) filter is used to increase the SNR of $r_c(k)$. The filter is designed to have a passband frequency of $0.1 \text{ Hz}$ and a stopband frequency of $15 \text{ Hz}$, which is considerably higher than the maximum Doppler spread ($\approx 4 \text{ Hz}$ for a speed of $0.5 \text{ m/s}$). Passband ripple of the filter is $0.05 \text{ dB}$ and it has $40 \text{ dB}$ attenuation at frequencies higher than $15 \text{ Hz}$.

\subsection{Deterministic RSS Models}\label{S:deterministic_models}

\begin{figure*}[!t]
\begin{centering}
\begin{tabular}{*2{>{\centering\arraybackslash}m{0.95\columnwidth}}}
\subfloat[Ellipse human model and its projection $P_{\omega}(x')$]{\includegraphics[height=2.2in]{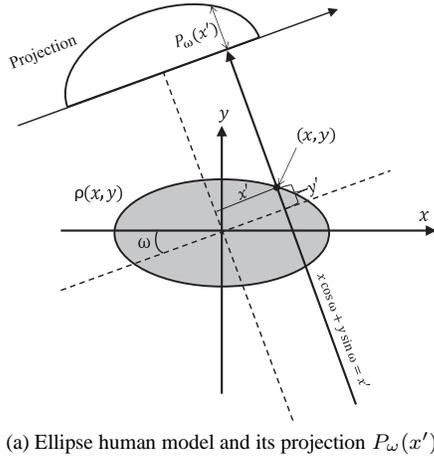}\label{fig:projection}} &
\subfloat[$r(k)$ vs. $\mathcal{S}(k, \boldsymbol{p_c})$ given in Eq.~\eqref{eq:shadowing_model}]{\includegraphics[width=\columnwidth]{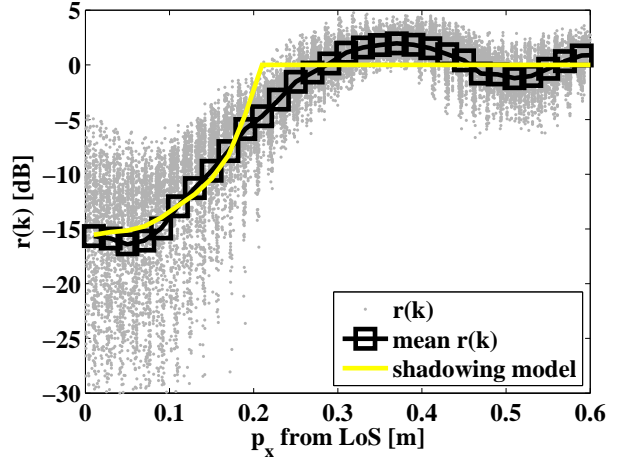}\label{fig:shadowing_model}}
\end{tabular}
\caption{Model for human-induced shadowing. In (b), $A=0.20 \text{ m}$, $B=0.20 \text{ m}$ and $\rho = 53 \text{ dB/m}$} 
\label{fig:shadowing}
\end{centering}
\end{figure*}

\subsubsection{Reflection Model}\label{S:reflection_model} 

A person moving in the close vicinity of the LoS can create additional multipath components by reflection \cite{patwari2011b,Liberti1996}. Considering the most simplistic scenario, a single multipath component caused by a \emph{single-bounce} reflection in addition to the LoS component as depicted in Fig.~\ref{fig:reflections}, then the channel gain in Eq.~\eqref{eq:channel} affecting the RSS measurements is given by
\begin{equation}\label{eq:reflected_signal}
	h(k) = \left|\alpha_{LoS}(k)e^{-j\phi_{LoS}} + \alpha_R(k) e^{-j\phi_R(k)}\right|^2.
\end{equation}
Under certain circumstances, the reflected component can be expressed with respect to the LoS component. First, the RX should be synchronized to the LoS component, $\phi_{LoS} = 0$. Second, the two components should experience similar path loss simplifying Eq.~\eqref{eq:reflected_signal} to
\begin{equation}
	h(k) = \left|\alpha_{LoS}(k)\left[1 + \Psi e^{-j\phi_R(k)}\right] \right|^2,
\end{equation}
where $0<\Psi<1$ describes the relation between $\alpha_{LoS}(k)$ and $\alpha_{R}(k)$. As given in Eq.~\eqref{eq:rss_mean_removed_filtered}, the mean is removed from the RSS in logarithmic scale which is equivalent to division in linear scale. Thus, the deterministic \emph{reflection model} is
\begin{equation}\label{eq:reflection_model}
\begin{aligned}
\mathcal{R}(k, \boldsymbol{p_c}) &= 10 \log_{10} \left(h(k)/\alpha_{LoS}^2(k)\right), \\
			       &= 10 \log_{10} \left(\Psi^2 + 2 \Psi\cos\phi_R(k)  +1 \right),
\end{aligned}
\end{equation}
where $\phi_R(k) = 2\pi\Delta_R(k)/\lambda_c$ and $\lambda_c$ is wave length of the carrier frequency. Excess path length of the reflected signal is defined as $\Delta_R = \lVert \boldsymbol{p}_{TX} - \boldsymbol{p}_{R}\rVert + \lVert \boldsymbol{p}_{RX} - \boldsymbol{p}_{R}\rVert - \lVert \boldsymbol{p}_{TX} - \boldsymbol{p}_{RX}\rVert$, where $\lVert \cdot \rVert$ is the Euclidean norm.

To simplify the notation, we exclude time dependence $k$ in the following equations and define the relation $\Psi$ between $\alpha_{LoS}$ and $\alpha_{R}$ as
\begin{equation}\label{eq:amplitude_proportionality}
	\Psi \triangleq  \psi_{0} \psi_{\bot} \left(\frac{d_{LoS}}{d_{LoS} + \Delta_R}\right)^{\eta/2},
\end{equation}
where $\psi_{0}$, $d_{LoS}$ and $\eta$ are experiment dependent and time-invariant reflection coefficient, LoS distance, and path loss coefficient in respective order. Further, $\psi_{\bot}$ is the time-variant Fresnel reflection coefficient \cite[pp. 114-125]{rappaport1996}
\begin{equation}\label{eq:reflection_coefficient}
\psi_{\bot} = \frac{\sin \theta_i- \sqrt{\varepsilon_r - \cos^2\theta_i}}{\sin \theta_i + \sqrt{\varepsilon_r - \cos^2\theta_i}}
\end{equation}
of perpendicular E-field polarization at the boundry of two dielectrics. In Eq.~\eqref{eq:reflection_coefficient}, $\varepsilon_r$ is relative permittivity whereas $\theta_i$ is the incidence angle of reflection as illustrated in Fig.~\ref{fig:reflections}.

There are a few observations about the reflection model defined in Eq.~\eqref{eq:reflection_model} and specifically parameter $\Psi$ defined in Eq.~\eqref{eq:amplitude_proportionality} that need to be elaborated. First, the reflection model is based on a single multipath component that reflects from point $\boldsymbol{p}_R$ with incidence angle $\theta_i$ that implicitly defines a tangent to the human ellipse model that minimizes $\Delta_R$. Excess path length $\Delta_R$ defines the additional path loss experienced by the reflected wave given in Eq.~\eqref{eq:amplitude_proportionality} as $\left(d_{LoS}/(d_{LoS} + \Delta_R)\right)^{\eta/2}$. Second, Fresnel reflection coefficient $\psi_{\bot}$ defines the energy preserved in the reflected wave. We assume that when a observable change in RSS is measured, $\theta_i$ is smaller than the \emph{Brewster angle} and therefore, parallel E-field polarization does not affect the RSS. Third, clothing is the dielectric boundary. For textiles, $\varepsilon_r$ is near $1.5$ and it does not vary much at $2.45 \text{ GHz}$ \cite{Sankaralingam2010} so that it can be assumed constant. Fourth, the empirical reflection coefficient $\psi_{0}$ is used to scale the reflection model to match the average measured change in RSS in reflection state. Thus, the reflection model is explicitly defined by $\boldsymbol{p_c}$, geometry and electric properties of the human ellipse model.

In Fig.~\ref{fig:reflection_model}, the changes in $r(k)$ with respect to $\Delta_R(k)$ are shown. For comparison, $\mathcal{R}(k, \boldsymbol{p_c})$ defined in Eq.~\eqref{eq:reflection_model} is illustrated. As shown, the empirical data closely resembles the model for reflections except at very small $\Delta_R(k)$ values where the temporal channel state changes to shadowing. In Section \ref{sec:parameter_sensitivity}, we investigate the effect of $\varepsilon_r$ and $\psi_{0}$ to the system performance.

\subsubsection{Shadowing Model}\label{S:shadowing_model}
	
\begin{figure*}[!t]
\begin{centering}
\begin{tabular}{ccc}
\mbox
{
\subfloat[Non-fading $\sim$ \emph{Log-normal}]{\includegraphics[width=\columnwidth*2/3]{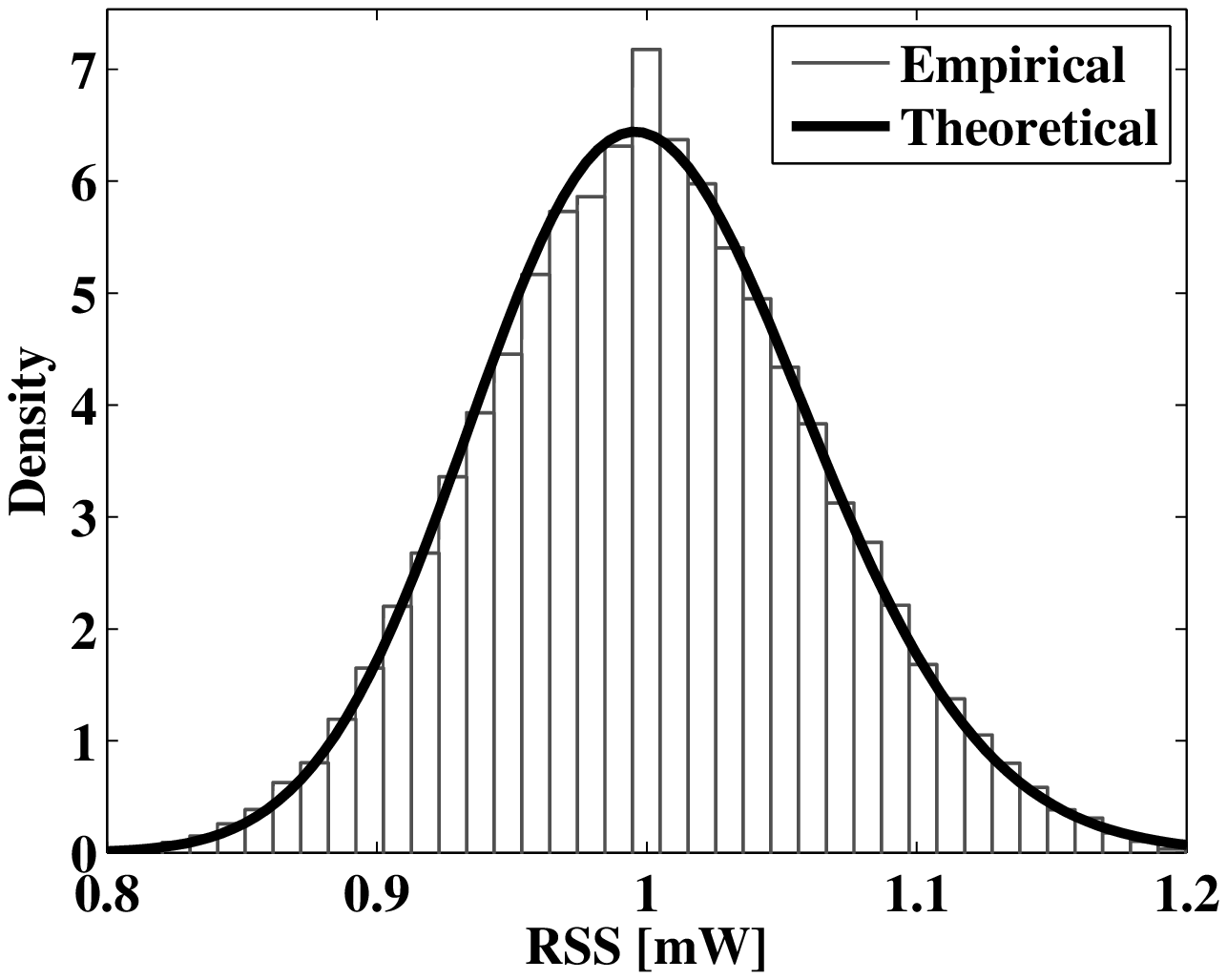}\label{fig:noise_distribution}}
\subfloat[Reflection $\sim$ \emph{Weibull}]{\includegraphics[width=\columnwidth*2/3]{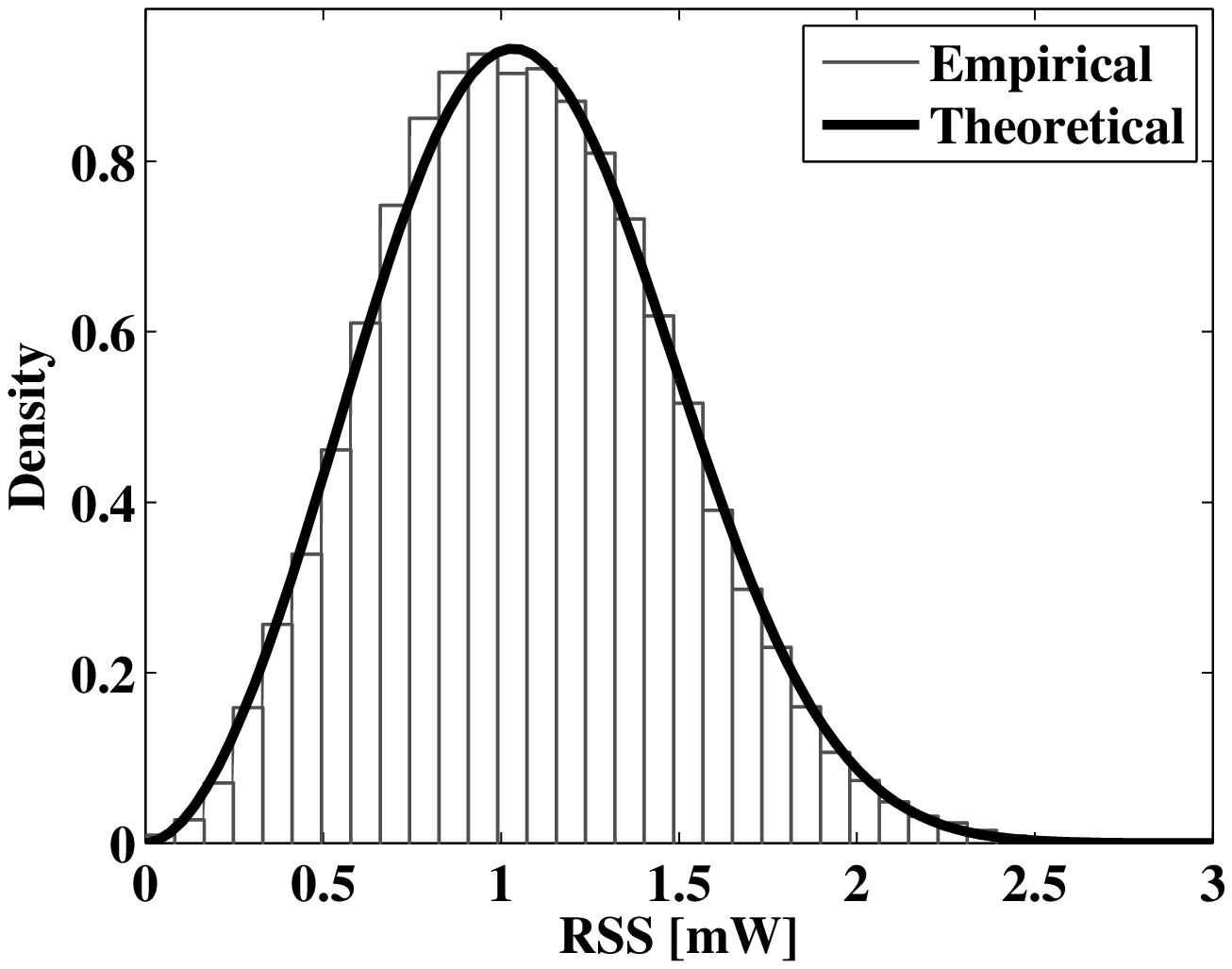}\label{fig:reflection_distribution}}
\subfloat[Shadowing $\sim$ \emph{Gamma}]{\includegraphics[width=\columnwidth*2/3]{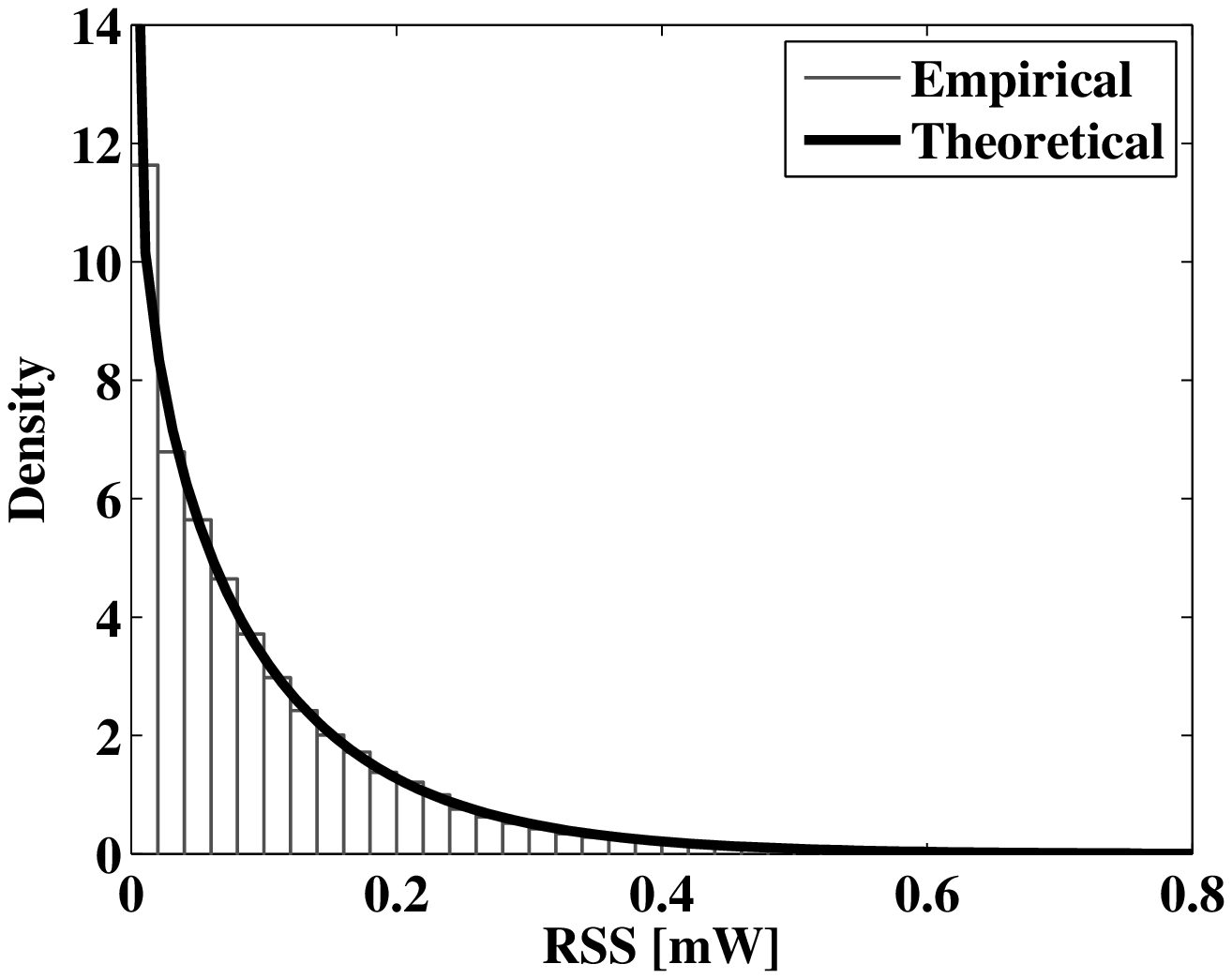}\label{fig:shadowing_distribution}}
}
\end{tabular}
\caption{Empirical and theoretical densities for the three temporal propagation channel states} 
\label{fig:distributions}
\end{centering}
\end{figure*}

RF signals can diffract, scatter, reflect and attenuate upon contact with the person making it a demanding task to accurately model human-induced RSS changes in shadowing state. However, the modeling effort can be considerably simplified by neglecting the other mechanisms and assuming transmission through the human body to contribute the most. In this case, attenuation can be represented by a \emph{line integral} of the attenuation field along a straight line from TX to RX as visualized in Fig.~\ref{fig:projection}.

The total attenuation along the line 
\begin{equation}\label{eq:line_equation}
	y' = x \cos\omega + y \sin\omega  - x' 
\end{equation}
caused by attenuation field $\rho(x,y)$ as illustrated in Fig.~\ref{fig:projection}, can be written as \cite[Ch. 3]{Kak1988}
\begin{equation}\label{eq:line_integral}
	P_{\omega}(x') = \int\limits_{-\infty}^{\infty} \int\limits_{-\infty}^{\infty} \rho(x,y) \delta(x \cos\omega + y \sin\omega - x') dx dy,
\end{equation}
where $\delta(\cdot)$ is Dirac delta function. In this paper, the cross section of a human is modeled as an ellipse with uniform electrical properties, i.e., $\rho(x,y) = \rho$. For such a geometry and properties, the closed form solution of Eq.~\eqref{eq:line_integral} is 
\begin{equation}\label{eq:projection}
    P_{\omega}(x')  =  \begin{cases}
                       \frac{2{\rho}AB}{a^2(\omega)}\sqrt{a^2(\omega) - (x')^2} & \text{if } \lvert x' \rvert \leq a(\omega)  \\
                       0 & \text{otherwise}
               \end{cases}
\end{equation}
where $A$ and $B$ are the semi-minor and semi-major axis of the ellipse and $a^2(\omega) = A^2\cos^2(\omega) + B^2\sin^2(\omega)$. The above formulations are closely related to the \emph{Radon transform} which is widely utilized in \emph{computerized tomographic imaging} (CTI) \cite[Ch. 3]{Kak1988}.

From the experiments, it is identified that the signals are attenuated more when the person is close to the transceivers. This finding can be explained through Fresnel zones which are concentric ellipsoids with radius \cite[pp. 126-135]{rappaport1996}
\begin{equation}\label{eq:freznel_radius}
	d_n = \sqrt{\frac{n\lambda_c d_{TX} d_{RX}}{d_{TX} + d_{RX}}},
\end{equation}
where $n$ is the Fresnel zone number, and $d_{TX} \text{ and } d_{RX}$ are the distances to the TX and RX in respective order. RF signals traversing through space have a greater flux density in $W/m^2$ the smaller the Fresnel radius is. Thus, the overall attenuation is relative to the area the person obstructs of this space. Since shadowing occurs only inside the first Fresnel zone ($n=1$), we approximate the relation by width of the person divided by the radius of the first Fresnel zone, i.e., $\kappa(p_y) = A/d_1$.

In this paper, the frame of reference is defined with respect to the TX-RX pair resulting that $\omega = 0$ and $x' = p_x$. Now, the measured decrease in RSS is equivalent to summing up the losses along the LoS. As a result, the deterministic \emph{shadowing model} can be expressed as
\begin{equation}\label{eq:shadowing_model}
	\mathcal{S}(k, \bm{p_c}) = -\kappa(p_y) P_{\omega}(p_x).
\end{equation}
In Fig.~\ref{fig:shadowing_model}, the RSS measurements with respect to $p_x$ are shown. For comparison, the analytical model for human-induced shadowing losses defined in Eq.~\eqref{eq:shadowing_model} is illustrated. As shown, the empirical data closely resembles the model for shadowing losses when the person is obstructing the LoS, i.e., $|p_x| \leq A$. In Section \ref{sec:parameter_sensitivity}, we investigate the effect of $A$ and $\rho$ to the system performance.

\subsection{Statistical RSS Models}\label{S:statistical_models}

In the following, the conditional densities of $r_c(k)$ in the three temporal propagation channel states are investigated. In this section, $r_c(k)$ are converted to linear scale using $r_{mW}(k;c) = 10^{r_c(k)/10}$ for convenience. The data are tested against various well known probability densities and the results are evaluated with the \emph{Kolmogorov-Smirnov} test \cite{massey1969} using a significance level of $5\%$. The tested distributions are: \emph{normal}, \emph{log-normal}, \emph{Rice}, \emph{Rayleigh}, \emph{Weibull}, and \emph{gamma}. The empirical and selected theoretical densities are shown in Fig.~\ref{fig:distributions}. Results of the statistical tests and parameters of the selected distributions are given in Table \ref{table:statistical_tests}. In the table, $\mathcal{H}_0$ depicts the \emph{null hypothesis} that the data is from the tested distribution and \emph{p-value} denotes the probability of the Kolmogorov-Smirnov statistic.


\subsubsection{Non-fading State}
For most of the time, the channel is time-invariant and RSS remains nearly constant \cite{bultitude1987} since the received signal only experiences a single realization of the fading process. Thus, it is expected that $r_{c}(k)$ follows the statistics of $\nu_c(k)$. For quadrature modulated communication systems such as the ones considered in this paper, the wideband noise is expected to follow the central chi-square distribution with two degrees of freedom. However, since the RSS is averaged over several periods of $s(t)$, $\nu_c(k)$ is composed of large number of independent noise components that are averaged. Due to the central limit theorem, $\nu_c(k)$ is expected to approach the Gaussian distribution yielding that $r_{mW}(k;c)$ in linear scale is log-normal for which the density function is given by
\begin{equation} \label{eq:noise_distribution}
     f_{r \vert s}\left(r \vert \mu, \sigma\right) = \frac{1}{r \cdot \sigma \sqrt{2\pi}}\exp \left(-\frac{(\ln (r) - \mu)^2}{2\sigma^2}\right).
\end{equation}
In Eq.~\eqref{eq:noise_distribution}, $\mu$ is the mean and $\sigma$ the standard deviation. Related works have also reported $\nu_c(k)$ to be Gaussian in logarithmic scale \cite{Zheng2012,Wilson2010,kaltiokallio2014}.

\subsubsection{Reflection State}

\begin{table}[!t]
    \caption{Statistical tests} 
        \centering 
	\renewcommand{\arraystretch}{1.15}
        \begin{tabular}{|c||c|c|c|}
	\hline\hline
	State & $s_1$ (Non-fading) & $s_2$ (Reflection) & $s_3$ (Shadowing) \\
	\hline
	Selected & \multirow{2}{*}{\emph{Log-normal}} & \multirow{2}{*}{\emph{Weibull}} & \multirow{2}{*}{\emph{Gamma}} \\
	Distribution & & &  \\
	\hline
	K-S & $\mathcal{H}_0:$ \emph{true} & $\mathcal{H}_0:$ \emph{true}  & $\mathcal{H}_0:$ \emph{true}  \\
	Results & \emph{p-val.} $:46.70\%$ & \emph{p-val.} $:54.75\%$ & \emph{p-val.} $:22.01\%$ \\
	\hline
	Estimated & $\mu=0$ & $a = 1.242$ & $a = 0.919$ \\
	Parameters & $\sigma=0.088$ & $b = 2.630$ & $b = 0.127$ \\
	\hline
        \end{tabular}
        \label{table:statistical_tests} 
\end{table}

As presented in Section \ref{S:reflection_model}, a person can create additional multipath components by reflection when they move in the close vicinity of the LoS. In this case, the amplitude of the reflected signal depends on the position, geometry, and electrical properties of the person, whereas time-delay $\tau_R(k)$ solely depends on $\Delta_R(k)$. Thus, in the considered scenario there are two clusters of multipath components with different amplitudes and arrival times. If the region is free of additional scatterers, the amplitude decay of the clusters is the same and therefore, the received signal power is obtained as the modulus sum of the multipath components raised to a certain exponent.

The described multipath propagation effect is known to yield Weibull distributed fading  \cite{Yacoub2002, Sagias2005} for which the density function with shape parameter $b$ and scale parameter $a$ is given by
\begin{equation}\label{eq:reflection_distribution}
f_{r \vert s}\left(r \vert a, b \right) = \frac{b}{a}\left( \frac{r}{a}\right)^{b-1}\exp \left(-\left( \frac{r}{a}\right)^b\right).
\end{equation}
In case the envelope follows a Weibull distribution, the signal power also follows the Weibull distribution with shape parameter $b/2$ \cite[p. 26]{Simon2005}.

\subsubsection{Shadowing State}

As electromagnetic waves propagate through a person, they are reflected, diffracted and scattered multiples of times due to differences in electrical properties of human tissue. However, most tissues' are comparable to or less than the wavelength in size resulting that the path length traveled by different waves is approximately equal. Consequently, the phase variation in the received signal is negligible and it is possible to argue that the phase is deterministic. Thus, the observed effect is solely on the amplitude so that power of the received signal is significantly reduced.
 
One approach to model amplitude fading in shadowing state is to consider the power transmitted through a person to be mostly scattered by human tissue. In this case, the electromagnetic waves are scattered by a large number of irregular surfaces (tissue). As a consequence, the received power intensity profile is expected to follow a gamma variate \cite{Jakeman1982}. Similar arguments hold for slow-fading in mobile propagation channels where the gamma distribution has been proposed to characterize shadow fading \cite{Abdi1999}. Therefore, $r_{mW}(k;c)$ is assumed to follow the gamma distribution with shape parameter $b$ and scale parameter $a$
\begin{equation} \label{eq:fading_distribution}
     f_{r \vert s}\left(r \vert a, b\right) = \frac{r^{b-1}}{a^b\Gamma(b)}\exp \left(-\frac{r}{a}\right), 
\end{equation}
where $\Gamma(\cdot)$ is the gamma function.

\section{Application} \label{S:application}
In this section, a simplistic DFL application is designed to make an accurate evaluation of the proposed models.  
The system consists of one transmitter and two receivers that are deployed in a LoS indoor environment as shown in Fig.~\ref{fig:experimental_setup}. The aim is to locate and track a single person as they enter and travel through the monitored area.  Tracking multiple people is not within the scope of this work and the readers are referred to \cite{bocca2013b} and the references therein for available solutions. The introduced system could for example be used to monitor people flow in the deployment area. Further, deploying multiple independent systems to corridors and doorways of a building would enable monitoring people flow throughout the building.

The localization algorithm is composed of two main tasks: first, estimating the temporal state of the propagation channel; second, estimating the person's location in case human-induced temporal fading is observed. The unobservable temporal state of the wireless medium is estimated with a hidden Markov model (HMM) and the statistical models introduced in Section \ref{S:statistical_models}. Respectively, the person's location is estimated using a particle filter and the deterministic models presented in Section \ref{S:deterministic_models}. The dynamics of the person are represented using a constant velocity model. In the subsequent sections, we refer to the temporal propagation channel state as \emph{link state} $s$, whereas to the location and velocity of the person as \emph{kinematic state} $\bm{x}$. The pseudo-code of the application is presented in Algorithm $1$.

\begin{center}
\renewcommand{\arraystretch}{1.2}
  \begin{tabular}{ l }
    \hline\hline
	\bf{Algorithm 1: } \it{Application} \\
    \hline
	\it{At time k, use HMM to estimate link state $\hat{s}(k)$} \\
	\bf{if} \it{particle filter is initialized} \bf{do} \\
		\quad \bf{if} $\hat{s}(k) = s_1 \; \text{ \it{ for all links}}$ \bf{do} \it{stop tracking} \\
		\quad \bf{else} \it{estimate kinematic state} $\bm{\hat{x}}_k$ \it{using} \bf{Algorithm 2}\\
	\bf{elseif} $\hat{s}(k) = s_3 \text{ \it{ for any link}}$ \bf{do} \it{initialize particle filter} \\
	\hline
  \end{tabular}
\end{center}

\subsection{Estimating the Link State}

In this paper, it is assumed that the current state of the propagation channel depends only on the previous state so that the system can be represented using a Markov chain. However, the state of the channel is not directly observable, i.e., it is not known which propagation mechanism is dominating at each given time instant. For this reason, the system is represented using a hidden Markov model (HMM) and the probability of each state is estimated using the RSS measurements and the statistical models.

The HMM calculates the probability of being in link state $s_i$ using the current measurement $r_{mW}(k)$, the state transition matrix $S$, the conditional densities of the observations $f_{r \vert s}$ and the initial state probability $f_0$. Then, the \emph{forward procedure} \cite[pp. 109-114]{Therrien1992} can be used to estimate the current state probabilities using
\begin{equation} \label{eq:forward_procedure}
     f_{s_i}(r_{mW}(k) \vert \xi) = \sum_{i=1}^{Q}\gamma_i(k),
\end{equation}
where $\xi = (S, \; f_{r \vert s}, \;f_0)$ denotes the HMM parameter set, $\gamma_i(k)$ the forward variable and $Q$ the number of states. At time instant $k$, $\gamma_i(k)$ can be calculated recursively using
\begin{equation} \label{eq:forward_variable}
     \gamma_i(k) = \left[\sum_{j=1}^{Q}\gamma_j(k-1) \cdot S_{i \lvert j}\right]f_{r \vert s}(r_{mW}(k)\lvert s_i),
\end{equation}
where $\gamma_i(1) = f_0 \cdot f_{r \vert s}(r_{mW}(1)\lvert s_i)$. Finally, the link state with the highest probability can be taken as the estimate. We use the maximum \emph{a posteriori} estimate defined as 
\begin{equation} \label{eq:state_estimate}
\hat{s}(k) = \argmax_i f_{s_i}(r_{mW}(k) \vert \xi)
\end{equation}
to determine the temporal state of the channel.

For the considered system, $Q=3$, $f_{r \vert s}$ are the densities derived in Section \ref{S:statistical_models}, and $S$ and $f_0$ can be assumed to be known \emph{a priori}, allowing one to utilize the HMM for estimating different temporal aspects of the system. In this paper, $\hat{s}(k)$ is used for estimating the LoS crossing instances and for determining when to start and stop tracking the person. Even though we do not demonstrate the effect of $S$ to the system performance, the application is robust to changes in $S$, as the system presented in \cite{kaltiokallio2014}. Since $f_{r \vert s}$ characterizes the RSS accurately, $\hat{s}(k)$ is the correct state as long as the transition probabilities satisfy $S_{i \lvert i} \gg S_{i \lvert j} \; \forall  \; i \neq j$. In Section \ref{S:results}, the following values are used
\begin{equation*}
S = \left[\begin{matrix} 0.95 & 0.05 & 0 \\ 0.025 & 0.95 & 0.025 \\ 0 & 0.05 & 0.95 \end{matrix}\right] \negthickspace\text{,}\quad
f_0 = \left[\begin{matrix} 0.7&0.2&0.1  \end{matrix}\right] \negthickspace\text{.}
\end{equation*}

\subsection{Localization and Tracking}\label{sec:tracking}

For the DFL application, we are interested in the location and trajectory of the person in an inertial frame of reference. In tracking applications, one of the widely utilized kinematic system models is the discrete-time constant velocity model \cite[Ch. 6]{BarShalom2001} for which the state-space representation can be written as
\begin{subequations} \label{eq:state_model}
\begin{align}
	\bm{x}(k+1) = \bm{F} \bm{x}(k) + \bm{G} \bm{w}(k) \\
	\bm{r}(k) = \bm{g}(k,\bm{x}(k)) + \bm{\nu}(k)
\end{align}
\end{subequations}
where $\bm{r}(k)$ is defined in Eq.~\eqref{eq:measurementmodel}, $\bm{w}(k)$ is zero-mean Gaussian process noise and 
$$
\bm{x}_k = \left[\begin{matrix} p_x(k) \\ v_x(k) \\ p_y(k) \\ v_y(k) \end{matrix}\right] \negthickspace\text{,}\quad
\bm{F} = \left[\begin{matrix} 1&T_s&0&0 \\ 0&1&0&0 \\ 0&0&1&T_s \\ 0&0&0&1 \end{matrix}\right] \negthickspace\text{,}\quad
\bm{G} = \left[\begin{matrix} \frac{1}{2}T_s^2 \\ T_s \\ \frac{1}{2}T_s^2 \\ T_s\end{matrix}\right] \negthickspace\text{.}\quad
$$
Thus, the trajectory of the person is estimated using the linear kinematic state model and the non-linear observation models defined in Section \ref{S:deterministic_models}.

Since the observation models are non-linear and the measurement noise is non-Gaussian for the described system (see Section \ref{sec:spatial_model_error}), we implement a particle filter to track the movements of the person. Particle filters are based on point mass representation of probability densities and they are especially suitable for non-linear/non-Gaussian problems where optimal algorithms such as the Kalman filter fail \cite{Arulampalam2002}. Particle filters have been successfully used in DFL applications to track the movements of a person e.g. in \cite{Wilson2012,Zheng2012,li2011,Guo2013}. The implemented particle filter is summarized in Algorithm $2$.

Initial distribution of the particles has a significant impact on the tracking performance. In order to reduce the uncertainty in the initial particle locations, the tracking is started only when one of the links transfers from $s_2 \text{ to } s_3$ since the person's perpendicular distance to the LoS is approximately known at this time instant (see Section \ref{sec:link_crossing_results}). At initialization, the particles are set uniformly on the LoS of the TX-RX pair. However, to account for the geometrical extent of the person, x-coordinate of the particles are shifted by $A = 0.11 \text{ m}$ towards the direction the person is approaching from (this is estimated from the state information of the links). Further, we suppose that the person is moving toward the monitored area by initializing $v_y$ to $0 \text{ m/s}$ whereas we initialize $v_x$ uniformly in $(0,1] \text{ m/s}$. The tracking is stopped when all links are in non-fading state $s_1$.

\begin{center}
\renewcommand{\arraystretch}{1.2}
  \begin{tabular}{ l }
    \hline\hline 
	\bf{Algorithm 2: } \it{Particle Filter} \\
    \hline
Predict state $\bm{x}_i(k) = \bm{F}\bm{x}_i(k-1) + \bm{G}\bm{w}$ using \eqref{eq:state_model} \\ 
Calculate observation  $\bm{z}_i(k) = g(\bm{x}_i(k))$  using \eqref{eq:signal} \\ 
Weight update $w_i(k) \propto f(\bm{r}(k) - \bm{z}_i(k)) \text{, } f \sim \mathcal{N}(0,\sigma_p)$ \\ 
Normalize $\tilde w_i(k) = \frac{w_i(k)}{\sum_{i=1}^{N} w_i(k)}$ \text{so that $\sum_{i=1}^{N} \tilde{w}_i(k) = 1$ } \\ 
Resample $\lbrace \bm{x}_i(k), \tilde{w}_i(k) \rbrace \rightarrow  \lbrace \frac{1}{N}, \bm{x}_i(k) \rbrace$ \\ 
Estimate $\bm{\hat{x}}(k) = \frac{1}{N}\sum_{i=1}^{N} \bm{x}_i(k)$ as mean of particles\\ 
	\hline
  \end{tabular}
\end{center}

\section{Experimental Evaluation} \label{S:experiments}
\subsection{Experiments}

\begin{figure}[!t]
\begin{centering}
\includegraphics[width=\columnwidth]{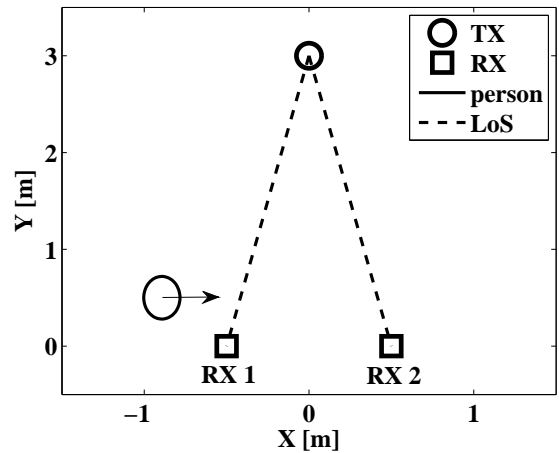}
\caption{Experimental setup} 
\label{fig:experimental_setup}
\end{centering}
\end{figure}

The performance of the system is demonstrated with a simplistic DFL system. For the purpose, a single transmitter and two receivers are deployed at the opposite walls of a corridor as shown in Fig.~\ref{fig:experimental_setup}. Overall, the system is tested in three different corridors each having different width, i.e., $2.0$, $3.0$ and $3.5$ m and they are labeled as \emph{Experiment} $1$, $2$ and $3$ in respective order. The receivers are placed a meter away from each other in the experiments. Controlled experiments are conducted in each corridor with a person moving at a known velocity ($v_x = \pm 0.5 \text{ m/s}$) along a path defined \emph{a priori} of deployment. The path intersects the LoS of the links multiple times at various y-coordinates. The used hardware and communication protocol are the same as the ones described in Section \ref{sec:measurements_collection}.

To retain the focus of the paper in the derived models, we do not present methods for learning the baseline RSS statistics or model parameters online. However, in practical applications online training is required and it is an important part of the overall system \cite{Kaltiokallio2012,Zheng2012,edelstein2013}. As a potential solution, we could exploit the method proposed in \cite{Kaltiokallio2012} to learn the reference RSS characteristics, adapt the online learning algorithms in \cite{Zheng2012} to update the RSS statistics, or use the background subtraction algorithm in \cite{edelstein2013} to calibrate the baseline RSS. On the other hand, the expectation maximization algorithm presented in \cite{li2011} could be used to estimate parameters of the spatial models online. In this paper, the experimental parameters are derived using the data collected in Section \ref{sec:measurements_collection} and the parameters are given in Table \ref{table:experimental_parameters}. Further, $A=0.2 \text{ m}$ for the shadowing model to minimize the modeling error for shadowing.

Even though the performance of the system is evaluated offline, the system is designed so that it is capable of online operation. Of the system components, the number of particles used by the filter contributes to the computational overhead the most. We set $N = 1000$ which results to an average computation time of $29 \text{ ms}$ per iteration using a standard laptop equipped with a $2.67$ GHz Intel Core i7-M620 processor and $8.0$ of GB of RAM memory. Because the computation time is smaller than sampling interval $T_s$, online operation is very possible.

\begin{table}[!t]
    \caption{Experimental parameters} 
        \centering 
        \begin{tabular}{c c l} 
        \hline\hline\ 
        Parameter & Value & Description \\
        \hline  
        $N$ & 1000  &  Number of particles \\ 
	$T_s$ & 0.032  &  Sampling interval per channel [s] \\ 
	$C$ & 16  &  Number of used channels \\ 
	$\bm{w}$ & [0.2 0.6] & Standard deviation of process noise \\ 
	$\sigma_p$ & 1.5 & Standard deviation of meas. noise [dB] \\ 
	$A$ & 0.11  & Semi-minor axis of human model [m] \\ 
	$B$ & 0.20 & Semi-major axis of human model [m]  \\ 
	$\psi_0$ & 0.5 & Reflection coefficient \\ 
	$\varepsilon_r$ & 1.5 & Relative permittivity \\ 
	$\eta$ & 2.0 & Path loss coefficient \\ 
        $\rho$ & 53.0 & Attenuation factor [dB/m] \\ 
	
        \hline 
        \end{tabular}
        \label{table:experimental_parameters} 
\end{table}

\subsection{Empirical Benchmark Models}

The spatial model introduced in Section \ref{S:deterministic_models} is compared to two empirical models commonly used in related literature. The tested models are the \emph{exponential model} presented by Li \emph{et. al} \cite{li2011} and the \emph{exponential-Rayleigh model} presented by Guo \emph{et. al} \cite{Guo2013}. The empirical parameters for both models are determined using the data collected in Section \ref{sec:measurements_collection} which were also used to validate the models presented in this paper.

In the exponential model, the RSS is modeled to decay exponentially with respect to the excess path length $\Delta$ and for the collected data, the model is given by
\begin{equation}\label{eq:exponential_fading_model}
g(k) = -16.77 e^{-\Delta/0.026},
\end{equation}
where $\Delta = \lVert \boldsymbol{p}_{TX} - \boldsymbol{p}_{c}\rVert + \lVert \boldsymbol{p}_{RX} - \boldsymbol{p}_{c}\rVert - \lVert \boldsymbol{p}_{TX} - \boldsymbol{p}_{RX}\rVert$ in which $\boldsymbol{p}_{c}$ is the center coordinates of the person. The exponential model only covers the RSS variations caused by shadowing and it was extended by Guo \emph{et. al} \cite{Guo2013} to account for reflections to some extent. The exponential-Rayleigh model consist of two exponential decays and for the collected data, the model is given by 
\begin{equation}\label{eq:exponential_rayleigh_fading_model}
g(k) = -15.77 e^{-\Delta/0.065} + 142.71 \Delta \cdot e^{-\Delta^2/0.010},
\end{equation}
where the first exponential decay captures the large losses caused by shadowing whereas the second captures the constructive fading effects caused by reflection when a person is in close proximity of the LoS. In the remainder of the paper, we refer to the system presented in this paper as \emph{three-state} model and the benchmark models as \emph{exponential} and \emph{exponential-Rayleigh} models.

\subsection{Validation Metrics}

In Section \ref{S:results}, the performance of the system is validated using the metrics defined below. The tracking accuracy is evaluated using mean absolute error (MAE) of the coordinate estimates 
\begin{equation} \label{eq:coordinate_error}
\begin{aligned}
	\bar{\epsilon}_x &= \frac{1}{K} \displaystyle\sum\nolimits_{k=1}^K \left\lvert p_x(k) - \hat{p}_x(k) \right\rvert, \\
	\bar{\epsilon}_y &= \frac{1}{K} \displaystyle\sum\nolimits_{k=1}^K \left\lvert p_y(k) - \hat{p}_y(k) \right\rvert,
\end{aligned}
\end{equation}
where $(p_x,p_y)$ depicts the true location, $(\hat{p}_x,\hat{p}_y)$ the estimated location and $K$ the total number of estimates. 

The performance of a tracking system based on a particle filter also requires a metric to indicate whether the posterior density has converged to the correct one or not. If the particles have converged to the correct density, it is expected that majority of the particles are located inside the human ellipse. Therefore, we also use the percentage of particles within the modeled human ellipse as an evaluation criteria. The ratio is defined as 
\begin{equation} \label{eq:ratio}
	\bar{\epsilon}_\% = \frac{ \sum_{k=1}^K N_{i \in \mathcal{A}}(k)}{K \cdot N} \cdot 100 \%,
\end{equation}
where $N_{i \in \mathcal{A}}(k)$ is the number of particles within area $\mathcal{A}$ spanned by the human ellipse model. In addition, we report the enhancement in accuracy in each experiment with respect to the least accurate system using 
\begin{equation} \label{eq:enhancement}
\bar{\epsilon}_R  = \frac{ \bar{\epsilon}_{\%} - \min \lbrace\bar{\epsilon}_{\%,j}\rbrace}{ \min \lbrace\bar{\epsilon}_{\%,j}\rbrace },
\end{equation}
where subscript $j = 1,2,3$ denotes the different models.

\section{Results} \label{S:results}
\subsection{Detecting Link Line Crossings}\label{sec:link_crossing_results}

\begin{figure}[!t]
\begin{centering}
\includegraphics[width=\columnwidth]{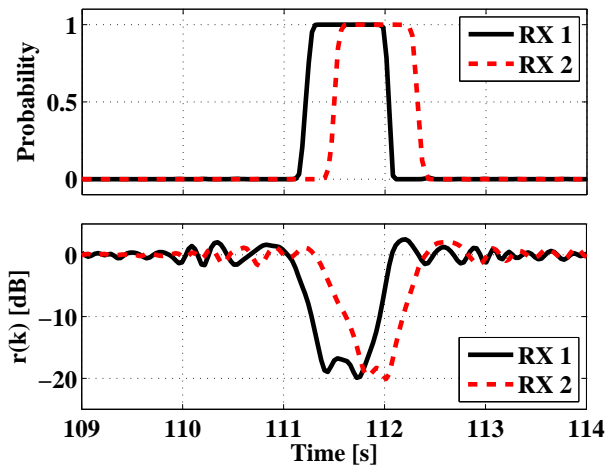}
\caption{Probability of being in shadowing state and $r(k)$ for a single link line crossing instance at $y = 2.5 \text{ m}$} 
\label{fig:hmm_state_estimate}
\end{centering}
\end{figure}

\begin{table*}[!t]
    \caption{Spatial model errors} 
        \centering 
	\renewcommand{\arraystretch}{1.1}
        \begin{tabular}{| l l l || c c c | c c c | c c c |} 
        
	\hline\hline
        \multicolumn{3}{|c||}{State}  & \multicolumn{3}{c|}{$s_1$ (\emph{non-fading})} 
				      &	\multicolumn{3}{c|}{$s_2$ (\emph{reflection})} 
				      & \multicolumn{3}{c|}{$s_3$ (\emph{shadowing})} \\
        \hline 
	
	{Model} & & & $\mathcal{H}_0$  & p-value & $\sigma_m$ 
				    & $\mathcal{H}_0$  & p-value & $\sigma_m$ 
				    & $\mathcal{H}_0$  & p-value & $\sigma_m$ \\
	\hline 
        \multicolumn{3}{|l||}{\emph{exponential}} & \emph{true} & $47.50$ & $0.363$  
				        & \emph{false} & $0$ & $1.666$
					& \emph{false} & $0$ & $5.223$ \\
					
	\multicolumn{3}{|l||}{\emph{exponential-Rayleigh}} & \emph{true} & $48.59$ & $0.363$  
				        & \emph{false} & $2.81$ & $1.533$
					& \emph{false} & $0$ & $5.115$ \\
					
	\multicolumn{3}{|l||}{\emph{three-state}} & \emph{true} & $47.50$ & $0.363$  
				        & \emph{true} & $25.02$ & $1.242$
					& \emph{false} & $0$ & $4.514$ \\
        \hline 
        \end{tabular}
        \label{table:model_errors} 
\end{table*}

The link state estimates $\hat{s}(k)$ of the HMM can be used in various ways and in this paper, they are used for detecting link line crossings and for estimating the temporal state of the propagation channel. In Fig.~\ref{fig:hmm_state_estimate}, the RSS measurements and probability of being in the shadowing state for a single link line crossing is shown when a person passes the LoS at $y = 2.5 \text{ m}$ during Experiment $2$. The link line crossing instances are estimated as the midpoint when the link is in shadowing state. These time instances are compared to the true crossing instances of the three experiments and on average, the difference is $72 \text{ ms}$ with a standard deviation of $33 \text{ ms}$. Taking into account the velocity of the person, the time difference corresponds to an error of $3.62 \text{ cm}$. Thus, it can be said that the system can detect the link line crossings with high accuracy because the error is considerably smaller than the person's geometric extent.

In Fig.~\ref{fig:hmm_state_estimate}, the person approaches the monitored area from the negative x-axis side and as shown, the sequence the receivers are in shadowing state is $RX 1 - RX 2$. From the link line crossing instances and the sequence information, the direction of movement can be easily estimated and in the experiments, it is correctly estimated at every link line crossing. Furthermore, if the y-coordinate would be known, the velocity of the person could be estimated from the difference between link line crossing instances. In this case, the average error of the velocity estimate is $2.83 \text{ cm/s}$ with a standard deviation of $3.18 \text{ cm/s}$. 

\subsection{Spatial Model Errors}\label{sec:spatial_model_error}

\begin{figure}[!t]
\begin{centering}
\includegraphics[width=\columnwidth]{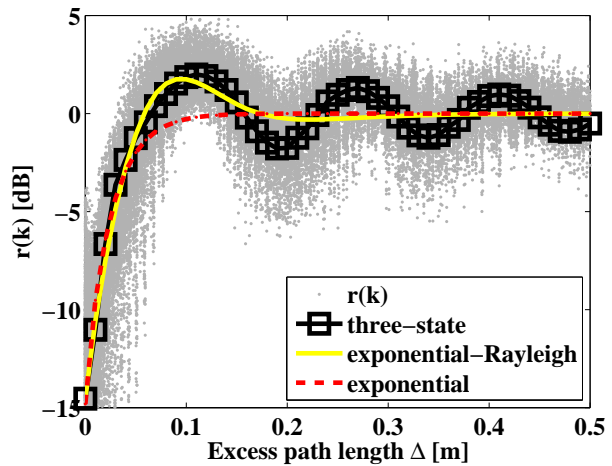}
\caption{$r(k)$ vs. the different models} 
\label{fig:model_comparison}
\end{centering}
\end{figure}

In Fig.~\ref{fig:model_comparison}, the different models with respect to $r(k)$ are shown. The exponential model was developed to model human-induced shadowing and as shown, when $\Delta < 0.06 \text{ m}$ the model corresponds to the RSS measurements closely. However, when $\Delta \geq 0.06 \text{ m}$ the model and measurements differ from one another. The exponential-Rayleigh model was developed to also account for human-induced temporal fading when the person is in the close proximity of the LoS. As shown in Fig.~\ref{fig:model_comparison}, the model is capable of capturing the first local maxima of the measurements. However, the exponential-Rayleigh model is incapable of capturing the subsequent local maxima’s and minima’s when $\Delta > 0.15 \text{ m}$. The shadowing model presented in this paper captures the large losses when the person is obstructing the LoS whereas the reflection model accounts for RSS variations when the person is moving in the close vicinity of the LoS. As illustrated, the three-state model and measurements correspond closely to each other.

In the following, the model error is evaluated using the Kolmogorov-Smirnov test and standard deviation of residual
\begin{equation} \label{eq:model_error}
	\sigma_m = \left( \frac{1}{K} \displaystyle\sum\nolimits_{k=1}^K \left( r(k) - g(k) \right)^2 \right)^{\frac{1}{2}}.
\end{equation}
The residual follow the joint statistics of modeling error and measurement noise. The models are accurate if the residual are zero-mean Gaussian since the measurement noise is assumed to be zero-mean Gaussian in Eq.~\eqref{eq:measurementmodel}. For evaluation, the Kolmogorov-Smirnov test with a significance level of $5\%$ is used to test \emph{null hypothesis} $\mathcal{H}_0$ that residual is normally distributed. Further, the \emph{p-values} of the Kolmogorov-Smirnov test as well as $\sigma_m$ are used for evaluation. The results are summarized in Table~\ref{table:model_errors} where the model error is evaluated separately in the three different link states. 

In non-fading state, the residual is Gaussian and all three models accept the null hypothesis as shown in the second column of Table~\ref{table:model_errors}. As expected, the results of the different models are close to one another since all of them estimate the measurements to be zero when the person is far away from the transceivers. In reflection state, the presented three-state model is the only one that accepts $\mathcal{H}_0$ as shown in the third column of Table~\ref{table:model_errors}. Furthermore, $\sigma_m$ is considerably lower than the values of the other two models because human-induced reflections are accounted for. Respectively, $\sigma_m$ of the exponential-Rayleigh model is lower than with the exponential model because it is able to capture the RSS changes caused by reflections when the person is very close to the LoS. In shadowing state, $\mathcal{H}_0$ is rejected for every model indicating that the residual is dominated by non-Gaussian modeling error. In addition, $\sigma_m$ values are significantly higher in shadowing state for the models revealing the difficulty in modeling human-induced shadowing accurately.

\begin{table*}[t]
    \caption{Tracking accuracy of the different models} 
        \centering 
	\renewcommand{\arraystretch}{1.2}
        \begin{tabular}{| l || >{\centering\arraybackslash}m{2.1cm} | >{\centering\arraybackslash}m{2.1cm} | 
	>{\centering\arraybackslash}m{1.1cm} | >{\centering\arraybackslash}m{1.1cm} |>{\centering\arraybackslash}m{2.1cm} |} 
	\hline\hline
	\multirow{2}{*}{Experiment and Model} & $\bar{\epsilon}_x \pm \sigma_x$ & $\bar{\epsilon}_y \pm \sigma_y$ &
					        $\bar{\epsilon}_\%$ & $\bar{\epsilon}_R$ & \emph{Sensitivity Region} \\
	
	\cline{2-6}
	& cm & cm & $\%$ & $\%$ & m$^2$\\
	\hline
	Ex. $1$ \emph{exponential} & $7.28 \pm 9.60$ & $15.77 \pm 20.23$ & $41.43$ & $0.00$ & $0.80$\\
	Ex. $1$ \emph{exponential-Rayleigh} & $6.44 \pm 9.90$ & $13.47 \pm 17.74$ & $50.11$ & $20.95$ & $1.33$\\
	Ex. $1$ \emph{three-state} & $\bm{3.16} \pm \bm{4.37}$ & $\bm{9.13} \pm \bm{12.96}$ 
				   & $\bm{69.43}$ & $\bm{67.58}$ & $\bm{2.95}$\\

	\hline			
	Ex. $2$ \emph{exponential} & $8.88 \pm 10.95$ & $27.40 \pm 31.32$ & $25.57$ & $0.00$ & $1.45$\\
	Ex. $2$ \emph{exponential-Rayleigh} & $5.17 \pm 6.60$ & $19.19 \pm 24.86$ & $45.74$ & $78.88$ & $2.38$\\
	Ex. $2$ \emph{three-state} & $\bm{2.50} \pm \bm{3.59}$ & $\bm{9.48} \pm \bm{16.66}$ 
				   & $\bm{78.11}$ & $\bm{205.48}$ & $\bm{4.96}$\\

	\hline			
	Ex. $3$ \emph{exponential} & $12.17 \pm 15.40$ & $36.84 \pm 34.59$ & $14.73$ & $0.00$ & $1.82$\\
	Ex. $3$ \emph{exponential-Rayleigh} & $8.33 \pm 10.52$ & $25.55 \pm 32.73$ & $30.62$ & $107.88$ & $2.97$\\
	Ex. $3$ \emph{three-state} & $\bm{4.00} \pm \bm{5.91}$ & $\bm{14.21} \pm \bm{23.91}$ 
				   & $\bm{66.97}$ & $\bm{354.65}$ & $\bm{6.08}$\\

	\hline
        \end{tabular}
        \label{table:tracking_accuracy} 
\end{table*}

\begin{figure*}[!t]
\begin{centering}
\begin{tabular}{ccc}
\mbox
{
\subfloat[\emph{exponential model}]{\includegraphics[width=\columnwidth*2/3]{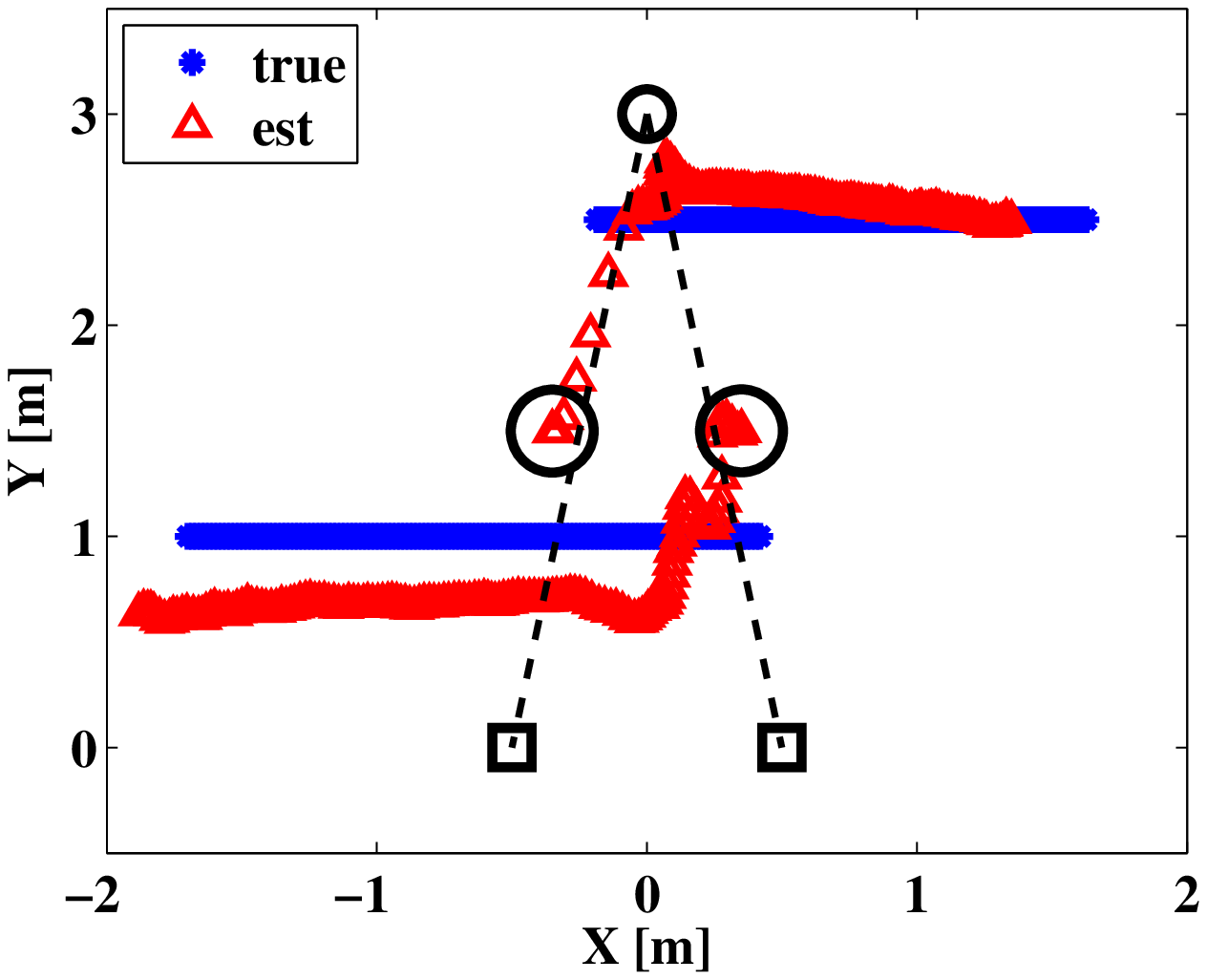}\label{fig:attenuation_trajectory}}
\subfloat[\emph{exponential-Rayleigh model}]{\includegraphics[width=\columnwidth*2/3]{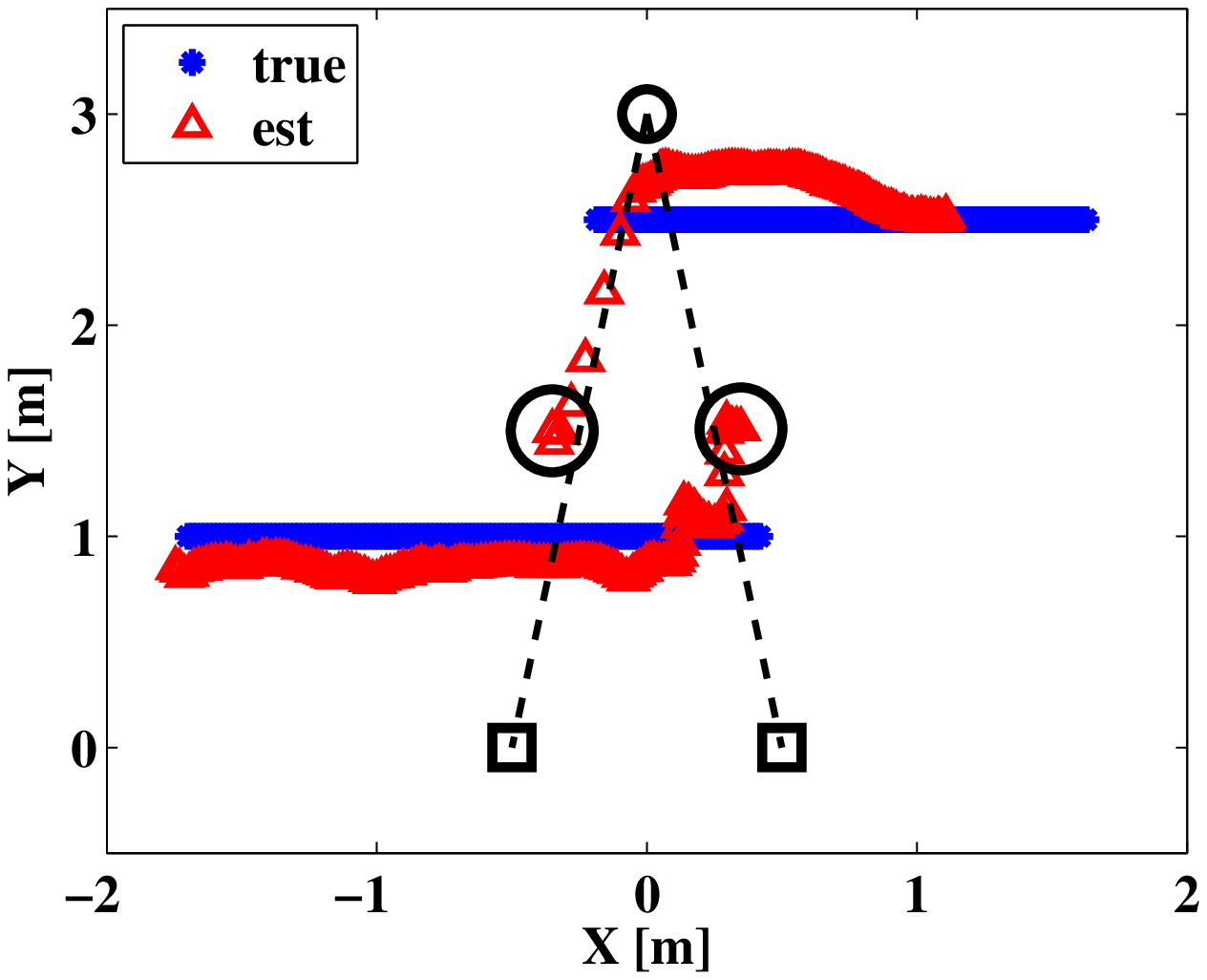}\label{fig:rayleigh_trajectory}}
\subfloat[\emph{three-state model}]{\includegraphics[width=\columnwidth*2/3]{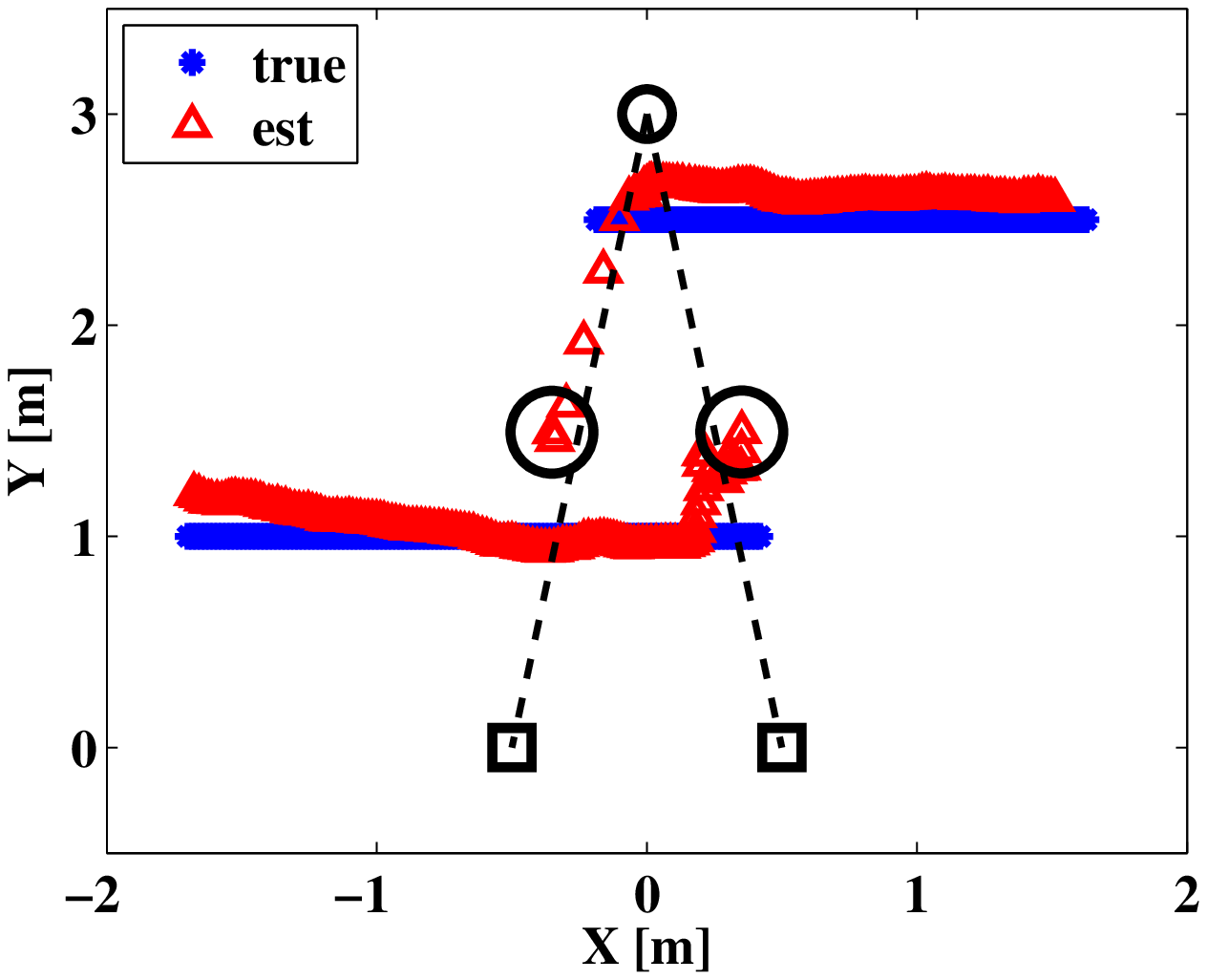}\label{fig:3state_trajectory}}
}
\end{tabular}
\caption{Example trajectories with the different models. The black circles illustrate mean of the initial particle cloud} 
\label{fig:trajectories}
\end{centering}
\end{figure*}

\subsection{Localization and Tracking}\label{sec:tracking_results}

To validate the tracking accuracy of the different models in the three experiments, $100$ Monte-Carlo simulations are performed with every model/experiment combination and in Table~\ref{table:tracking_accuracy}, the results are summarized. In Fig.~\ref{fig:trajectories}, two example trajectories in Experiment $2$ using the different models are shown. On average, the position estimates are more accurate using the three-state model than with the other two models and as the transceiver distance grows, the enhancement in performance increases. Thus, it is mandatory to take into account also human-induced reflections to develop more accurate DFL systems. The readers are invited to view the accompanying video that demonstrates the performance of the developed system and derived models \cite{link_line_monitoring_video}.

It is expected that the models are capable of tracking the x-coordinate of the person with considerably higher accuracy compared to the y-coordinate. Reason being, the models are generally more informative with respect to the coordinate that is perpendicular to the LoS of the TX-RX pair. Since the nodes are deployed in a corridor, the LoS of the links are almost parallel and therefore, the x-coordinate errors are only a fraction of the y-coordinate errors as shown in Table~\ref{table:tracking_accuracy}. 

On average, the accuracy of both coordinate estimates decays as the sensor distance is increased due to the additive uncertainty in the larger area to be monitored. However, the exponential-Rayleigh and three-state models are capable of dealing with this uncertainty better since they take into account the human-induced reflections. The exponential-Rayleigh model is able to capture the first constructive reflection ($\Delta_R = \lambda_c/2 \approx 6 \text{ cm}$), whereas the three-state model can also account for the destructive reflections located at $\Delta_R = n \cdot \lambda_c \text{, } n = 1,2,3,\cdots$ and constructive reflections located at $\Delta_R = n \cdot \lambda_c/2 \text{, } n = 1,3,5,\cdots$ as long as the amplitude of the reflected wave is large enough. Thus, the uncertainty using the three-state model is decreased even further which results to the best tracking accuracy.

\begin{figure*}[!t]
\begin{centering}
\begin{tabular}{ccc}
\mbox
{
\subfloat[\emph{Measurement noise}]{\includegraphics[width=\columnwidth*2/3]{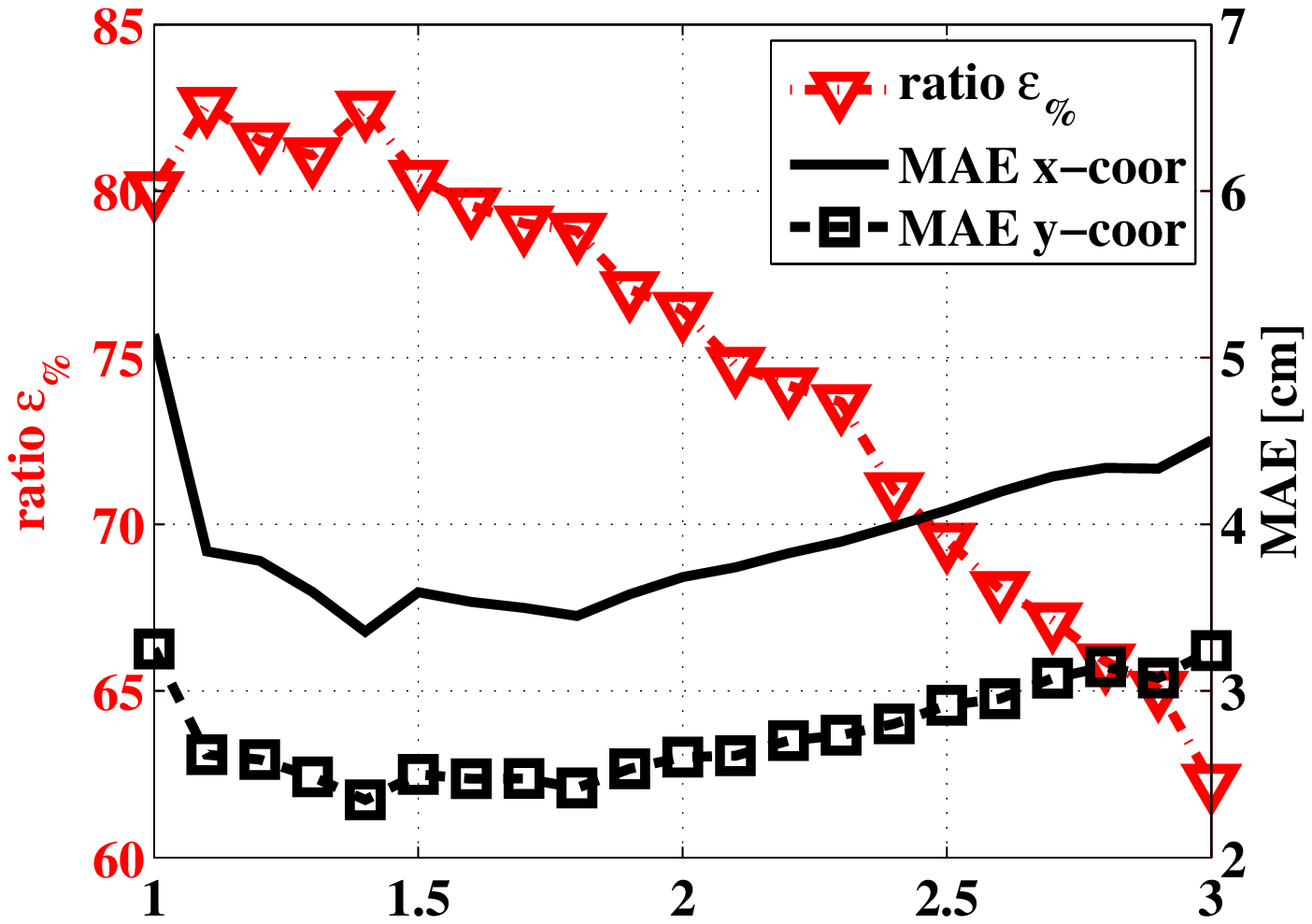}\label{fig:measurement_noise}}
\subfloat[\emph{Reflection model}]{\includegraphics[width=\columnwidth*2/3]{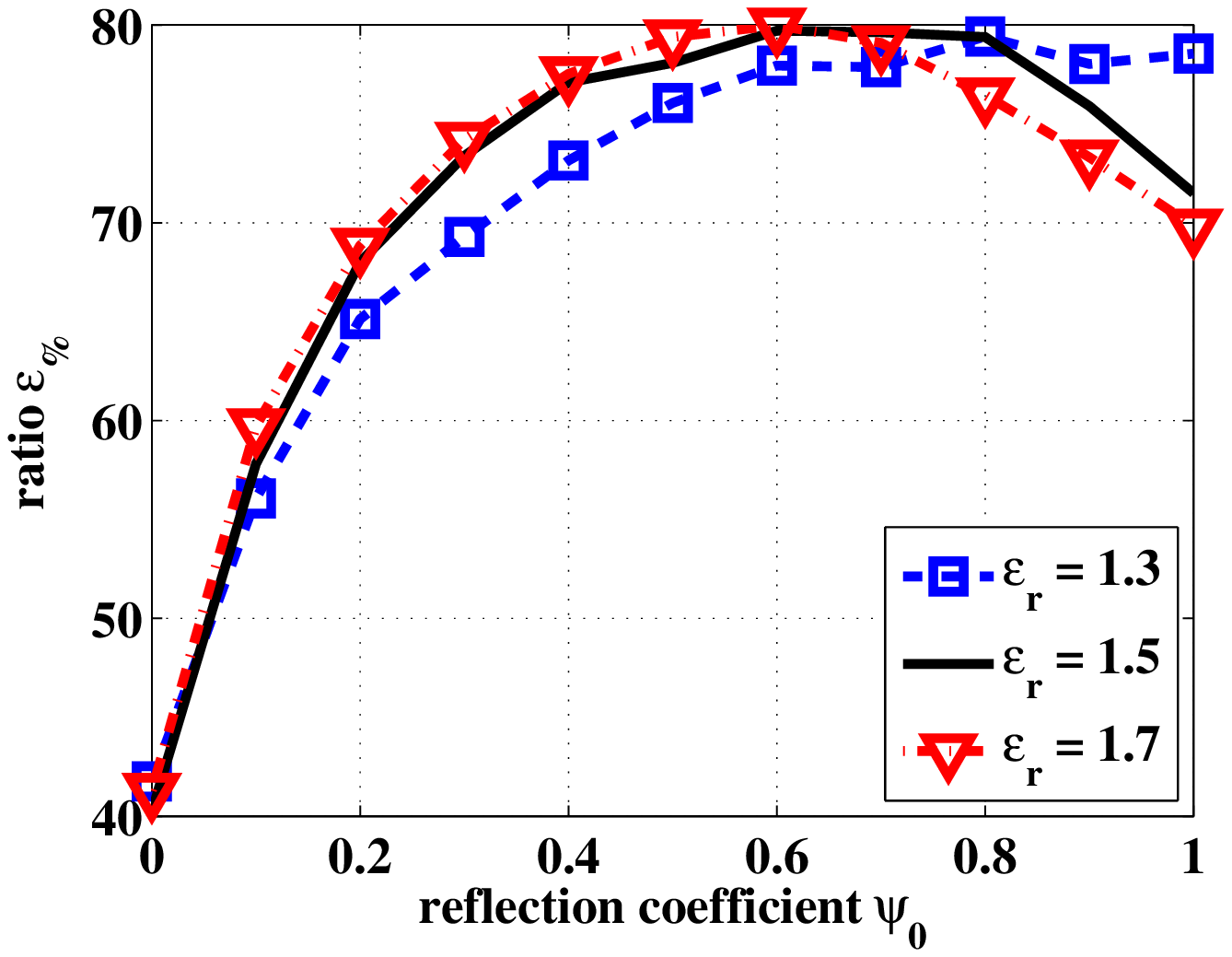}\label{fig:reflection_parameters}}
\subfloat[\emph{Shadowing model}]{\includegraphics[width=\columnwidth*2/3]{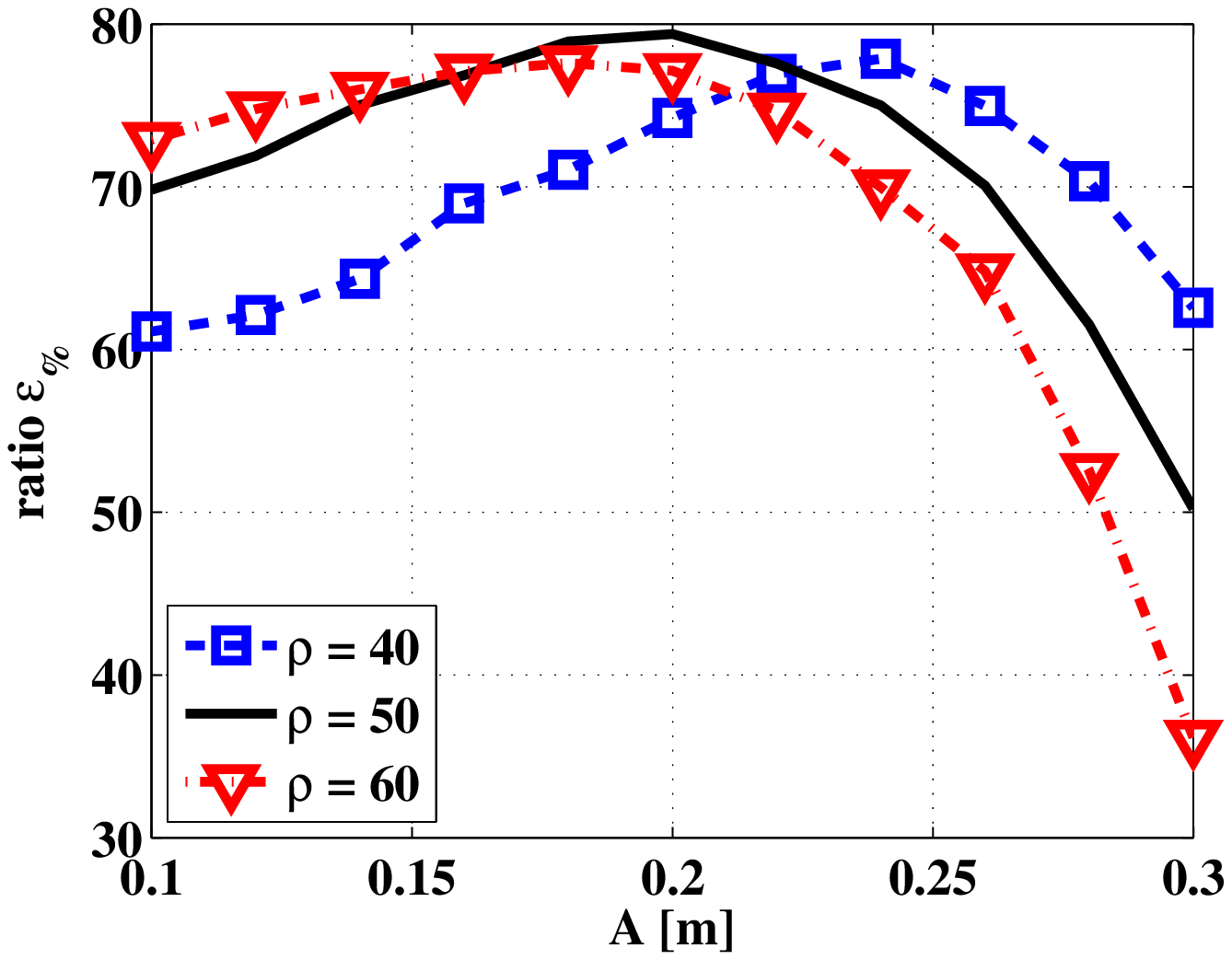}\label{fig:shadowing_parameters}}
}
\end{tabular}
\caption{Sensitivity of the DFL system to various parameter changes} 
\label{fig:parameter_sensitivity}
\end{centering}
\end{figure*}

In the last column of Table~\ref{table:tracking_accuracy}, the size of the sensitivity region for the different models is shown. The sensitivity region is determined by $\Delta$ and for the benchmark models, we search for the value of $\Delta$ where the models deviate more than $\pm 1 \text{ dB}$ from the mean $r(k)$. As an outcome, $\Delta = 0.06 \text{ m}$ for the exponential model, whereas $\Delta = 0.15 \text{ m}$ for the exponential-Rayleigh model. The value of $\Delta$ must be defined differently for the three-state model because $g(k)$ is within $\pm 1 \text{ dB}$ for all $\Delta$ values. However, resolution of the receiver's RSS output dictates whether the change in the propagation channel is observable or not. Thus, we use the link state estimates of the HMM and calculate the average $\Delta$ value when the HMM changes from $s_1 \text{ to } s_2$ or $s_2 \text{ to } s_1$, yielding $\Delta = 0.50 \text{ m}$ for the three-state model. As shown in Table~\ref{table:tracking_accuracy}, the sensitivity region of the three-state model is approximately three times larger than with the exponential model and twice the size with respect to the exponential-Rayleigh model. Thus, it is possible to achieve higher accuracy and to monitor a larger area with the three-state model.

The ratio of particles within the modeled human-ellipse, $\bar{\epsilon}_\%$, and the enhancement in performance, $\bar{\epsilon}_R$, are given in the fourth and fifth columns of Table~\ref{table:tracking_accuracy} in respective order. The trend of $\bar{\epsilon}_\%$ is the same as for the coordinate estimates. The three-state model outperforms the benchmark systems in every experiment and the enhancement in accuracy is greater at larger transceiver distances as indicated by $\bar{\epsilon}_R$. It is to be noted that $\bar{\epsilon}_\%$ can never achieve $100 \%$ accuracy due to uncertainty in the initial estimate. When the particle filter is initialized, most of the particles are outside the modeled human ellipse and as the filter converges closer to the true trajectory, more and more particles are within the ellipse.

\subsection{Parameter Sensitivity}\label{sec:parameter_sensitivity}

In the following, the effect of various parameters is investigated using Experiment $2$. The tested parameters are: $\sigma_p$ used in the weight update of the particle filter, reflection model parameters $\varepsilon_r$ and $\psi_0$, and shadowing model parameters $\rho$ and $A$. Even though results of the other two experiments are not presented, the trends of those curves are similar to the ones obtained with Experiment $2$.

In Fig.~\ref{fig:measurement_noise}, $\bar{\epsilon}_\%$, $\bar{\epsilon}_x$ and $\bar{\epsilon}_y$ illustrated as functions of measurement noise $\sigma_p$. On average, the ratio of particles inside the human model $\bar{\epsilon}_\%$ decreases when $\sigma_p$ is grown since the dispersion of the particle cloud is larger. However, selecting $\sigma_p$ too small typically leads to the situation where the posterior density converges to the incorrect one, resulting to a reduction in tracking accuracy as illustrated by $\bar{\epsilon}_x$ and $\bar{\epsilon}_y$ in Fig.~\ref{fig:measurement_noise}. Using Eq.~\eqref{eq:model_error} to calculate the standard deviation of measurement noise during Experiment $2$ results to $\sigma_p=1.44$. As shown, values near $\sigma_p=1.44$ result to a good tradeoff between accuracy and $\bar{\epsilon}_\%$.

The effect of reflection model parameters to system performance is shown in Fig.~\ref{fig:reflection_parameters}. For textiles, the relative permittivity is near $1.5$ and it does not vary much at $2.45 \text{ GHz}$ \cite{Sankaralingam2010}. As shown, small changes in $\varepsilon_r$ do not effect the results considerably and $\varepsilon_r = 1.5$ is a resonable value to be used with the reflection model. On the other hand, the system is notably more sensitive to the variations in $\psi_0$. The coefficient is used to scale the reflection model to match the average measured change in RSS. Values too small diminsh the model's effect on tracking whereas too large values result that the measurements and model do not correspond one another. Interestingly, when $\psi_0=0$ the reflection model has no impact on the system performance. Still, $\bar{\epsilon}_\% \approx 42 \%$, which is much higher than $\bar{\epsilon}_\% = 25.57 \%$ of the exponential model. Thus, the presented shadowing model is more informative than the exponential decay model.

In Fig.~\ref{fig:shadowing_parameters}, $\bar{\epsilon}_\%$ shown as a function of attenuation factor $\rho$ and $A$ of the human ellipse model. Increasing $A$ results to an enhancement in accuracy and selecting $A = 0.2 \text{ m}$ results to good performance despite the value of $\rho$. However, if $A$ is set too large, it starts to degrade the system performance because the model predicts the link to be shadowed whereas in reality, the system is experiencing constructive fading due to reflection. Furthermore, the reduction in performance is larger the higher the attenuation factor is.

\section{Conclusions} \label{S:conlcusions}
In this paper, a three-state temporal RSS model is presented and it is demonstrated that the measurements are dictated by electronic noise, reflection or shadowing; depending on the location of the person. Statistical and spatial models for the different states are derived and based on the models, a simplistic DFL application is developed with the aim to estimate the temporal state of the propagation channel and kinematic state of the person. Compared to empirical models presented in earlier works, the presented system achieves higher tracking accuracy while increasing the sensing region of the transceivers.

One of the key contributions of the paper is augmenting the vastly used time-varying two-state channel model with an additional state. In the proposed three-state model, human-induced temporal fading is divided into two states. In one of the states, a person is affecting the amplitude of the received signal by shadowing whereas in the other state, a person is mainly affecting the phase of non-dominant multipath components by reflection. The concepts, findings and models of this paper have a significant impact on DFL applications and in future work, use of the models in cluttered indoor environments will be studied to enhance the performance of existing RF sensor network applications.

\end{document}